%% file: JCP_2018.tex
\documentclass[preprint,11pt,3p]{elsarticle}

\usepackage[T1]{fontenc}
\usepackage[utf8]{inputenc}
\usepackage{amssymb}
\usepackage{amsmath}
\usepackage{latexsym}
\usepackage{mathtools}
\usepackage[format=hang,font=small,labelfont=bf]{caption}
\usepackage[outdir=./figures/]{epstopdf}
\usepackage{graphicx}
\usepackage{subfigure}
\usepackage{wasysym}
\usepackage[dvipsnames]{xcolor}
\usepackage{multirow}
\usepackage{units}
\usepackage{tikz}
\usepackage[toc,page]{appendix}
\usepackage{natbib}
\usepackage{booktabs}
\usepackage{multirow}
\usetikzlibrary{arrows,intersections}

\graphicspath{{figures/}}

\include{preamble}

\journal{Journal of Computational Physics}

\begin{document}

\begin{frontmatter}

\title{Numerically stable formulations of convective terms for turbulent compressible flows
}

\author[label1]{G. Coppola}
\address[label1]{Universit\`a di Napoli ``Federico II'', Dipartimento di Ingegneria Industriale, Napoli, Italy}
\ead{gcoppola@unina.it}

\author[label1]{F. Capuano}
\author[label2]{S. Pirozzoli}
\address[label2]{Universit\`a degli Studi di Roma ``La Sapienza'', Dipartimento di Meccanica e Aeronautica, Roma, Italy}

\author[label1]{L. de Luca}

 \begin{abstract}
A systematic analysis of the discrete conservation properties of 
non-dissipative, central-difference approximations of the compressible 
Navier-Stokes equations is reported. 
A general triple splitting of the nonlinear convective terms is considered, 
and energy-preserving formulations are fully characterized by deriving 
a two-parameter family of split forms.
Previously developed formulations reported in literature are shown to be 
particular members of this family; novel splittings are introduced and discussed
as well. 
Furthermore, the conservation properties yielded by different choices for the 
energy equation (i.e. total and internal energy, entropy) are analyzed
thoroughly.
It is shown that additional preserved quantities can be obtained through
a suitable adaptive selection of the split form within the derived family.
Local conservation of primary invariants, which is a fundamental 
property to build high-fidelity shock-capturing methods, is also discussed in the paper. 
Numerical tests performed for the Taylor-Green Vortex at zero viscosity fully confirm the 
theoretical findings, and show that a careful choice of both the splitting and the energy
formulation can provide remarkably robust and accurate results.
\end{abstract}


\end{frontmatter}
\section{Introduction} \label{sec:Intro}

It is well known that standard central finite-difference approximations
of the equations governing fluid flow are susceptible to numerical nonlinear 
instability when used at or near zero viscosity,
owing to accumulation of aliasing errors resulting from discrete evaluation of the 
convective terms~\citep{Phillips1959}.
Such shortcoming has also been traced to failure
in discretely preserving the secondary quadratic invariants associated with the conservation
equations~\citep{lilly_65,PerotAR}.
For instance, total kinetic energy is conserved from the Euler equations in the
incompressible limit in unbounded or periodic domains, and failure
to discretely satisfy this property typically leads to flow divergence. 
Several attempts have been made over the years to develop numerical
methods to solve the incompressible and the compressible flow equations which replicate
quadratic conservation properties in discrete sense,
both in finite differences and in finite-volume 
frameworks~\citep{Jameson2008,Kok2009,RozemaJT2014,Coppola_AIMETA2017}.
Most efforts made so far loosely rely on the idea of expanding the
convective derivatives to `skew-symmetric' split form, with goal objective of either minimizing the
aliasing error~\citep{Kravchenko1997,Blaisdell1996},
or to discretely preserve total kinetic energy~\citep{Feiereisen1981,Morinishi1998,Morinishi2010}.
Additional numerical robustness in the presence of density variations was gained 
by \citet{Kennedy2008} through the use of `triple splitting' of the convective terms.
Although splitting of the convective derivatives may guarantee good numerical stability properties,
the resulting discrete approximations are not generally expressible in
locally conservative form, i.e. a numerical flux cannot always be defined, hence
primary conservation properties may be violated, which is especially concerning
if the scheme has to be used as a building block of hybrid shock-capturing algorithms~\citep{JohnsenJCP2010}.
\citet{Pirozzoli2010} showed that locally conservative formulations with arbitrary order of accuracy are possible
for some types of splittings, and showed that a particular member of the family of splittings 
introduced by \citet{Kennedy2008} yields discrete energy preservation, and it is particularly stable. 

Discretization of the energy equation is also known to be a sensitive issue in the numerical
solution of the compressible Navier-Stokes equations. Of special importance is the choice of the 
form of energy equation among analytically equivalent ones, namely the internal energy, 
the total energy, the entropy equation and so on.
For instance, good numerical stability properties were found by~\citet{Honein2004} when using the entropy equation.
When the total energy equation is used, several arrangements are possible for the convective 
and pressure fluxes~\citep{Blaisdell1996,Kennedy2008,Pirozzoli2010}, which may 
lead or not to consistency with the internal and the kinetic energy equation, 
hence leading to different numerical stability properties.

It is the main goal of this paper to present a systematic study of the numerical conservation properties
of several central-difference approximations applied to the compressible Navier-Stokes equations.
First, in Sections~\ref{sec:ProbForm}, \ref{Sec:Splitting} we focus on the issue of splitting of the convective terms, and show that a more general family 
of energy-conserving splittings of the mass and momentum equations exists than currently known, 
and some of which also lead to local conservation of the primary invariants.
The properties of members of this wide family are presented and discussed, also in terms of empirically testing 
their numerical robustness.
Second, in Section~\ref{Sec:EnergyEq} we focus on the issue of the most appropriate formulation for the energy equation.
For that purpose, we consider formulations including the total and internal energy, and the entropy,
and bring out discrete conservations properties ensuing from different splittings. 
We also consider `dynamic' splittings for the energy equation(s), whereby additional conservation
properties are obtained through suitable adaptive selection of a free parameter in the family of available 
energy-conserving splittings. Numerical experiments are finally presented in Section~\ref{Sec:Tests} to support the validity of the theoretical inferences.

\section{Problem formulation} \label{sec:ProbForm}

The Navier-Stokes equations for a compressible flow can be written as
\begin{align}
\dfrac{\partial \rho}{\partial t} &= -\dfrac{\partial \rho u_j}{\partial x_j} \;, \label{eq:mass} \\[3pt]
\dfrac{\partial \rho u_i}{\partial t} &= -\dfrac{\partial \rho u_ju_i}{\partial x_j} -\dfrac{\partial p}{\partial x_i} + \dfrac{\partial \tau_{ij}}{\partial x_j} \;, \label{eq:momentum} \\[3pt]
\dfrac{\partial \rho E}{\partial t} &=  -\dfrac{\partial \rho u_jE}{\partial x_j} -\dfrac{\partial p u_j}{\partial x_j} + \dfrac{\partial \tau_{ij}u_i}{\partial x_j}
+ \dfrac{\partial }{\partial x_j} \left( k_c\dfrac{\partial T}{\partial x_j} \right) \;, \label{eq:energy}
\end{align}
where $E=u_i u_i/2 + c_v T$ is the total energy per unit mass and $\tau_{ij}$ is the stress tensor, 
for which the usual relation is assumed
\begin{equation*}
\tau_{ij} = \mu \left( \dfrac{\partial u_i}{\partial x_j} + \dfrac{\partial u_j}{\partial x_i} \right) - \dfrac{2}{3} \mu \dfrac{\partial u_k}{\partial x_k}\delta_{ij}.
\end{equation*}
The standard is here adopted whereby $\rho$ is the density, 
$u_i$ are the cartesian components of the velocity field, $p$ is the pressure, 
$T$ is the temperature, $k$ is the thermal conductivity, $c_v$ the specific 
heat at constant volume and $\mu$ is the molecular viscosity. 
Closure of the system is achieved by the ideal equation of state 
$p = \rho R T$, with $R$ the universal gas constant. 
Equations \eqref{eq:mass}-\eqref{eq:energy} represent the viscous 
balance of mass, momentum and total energy, and fully describe 
the motion of a compressible viscous fluid, once the equation of state 
and a suitable dependence of $\mu$ with temperature has been specified.

By introducing the internal energy $e = c_vT$
and the entropy  $s=c_v\ln(p/\rho^{\gamma})$, with $\gamma=c_p/c_v$
and $c_p$ the specific heat at constant pressure, the following induced balance 
equations are easily derived
\beq
\dfrac{\partial \rho e}{\partial t} =
-\dfrac{\partial \rho u_je}{\partial x_j}
-p\dfrac{\partial u_j}{\partial x_j} +
\tau_{ij}\dfrac{\partial u_i}{\partial x_j}+
\dfrac{\partial }{\partial x_j} \left( k\dfrac{\partial T}{\partial x_j} \right),
\label{eq:intenergy}
\eeq
\beq
\dfrac{\partial \rho s}{\partial t} =
-\dfrac{\partial \rho u_js}{\partial x_j}
+\dfrac{1}{T}\left[
\tau_{ij}\dfrac{\partial u_i}{\partial x_j}+
\dfrac{\partial }{\partial x_j} \left( k\dfrac{\partial T}{\partial x_j} \right)\right] \;,
\label{eq:entropy}
\eeq
any of which can be employed in place of the total energy, Eq.~\eqref{eq:energy}, 
to fully describe the motion and the state of the fluid.

The convective terms in Eqs.~\eqref{eq:mass}-\eqref{eq:entropy} have a common 
structure which can be summarized as
\begin{equation}
\mC=\dfrac{\partial \rho u_j \phi}{\partial x_j}, \label{eq:convgen}
\end{equation}
where $\phi$ equals unity, $u_i$, $E$, $e$ and $s$ for the mass, momentum, total 
and internal energy and entropy, respectively.
Integration of the governing equations readily shows that convective terms preserve the 
total amount of any conserved quantity. 
The associated invariants are hereafter referred to as linear invariants.
By applying the standard product rule, the generic convective term can 
be written in different analytically equivalent forms. 
Due to the cubic nonlinearity, there are five basic forms in which it can be expressed
\begin{align}
\mC^D &= \dfrac{\partial \rho u_j \phi }{\partial x_j}, \label{eq:CD}\\
\mC^{\phi} &= \phi \dfrac{\partial \rho u_j}{\partial x_j} + \rho u_j\dfrac{\partial \phi}{\partial x_j}, \label{eq:CF}\\
\mC^{u} &= u_j\dfrac{\partial \rho \phi}{\partial x_j}+\rho \phi \dfrac{\partial u_j}{\partial x_j}, \label{eq:CB}\\
\mC^{\rho} &= \rho \dfrac{\partial u_j \phi}{\partial x_j} + \phi u_j\dfrac{\partial \rho}{\partial x_j}, \label{eq:CK}\\
\mC^L &= \rho \phi \dfrac{\partial u_j}{\partial x_j} + \rho u_j\dfrac{\partial \phi}{\partial x_j} + \phi u_j\dfrac{\partial \rho}{\partial x_j}. \label{eq:CL}
\end{align}
Equation~\eqref{eq:CD} is the usual divergence form, 
whereas Eqs.~\eqref{eq:CF} and \eqref{eq:CB}
were firstly used in conjunction with $\mC^D$ by
\citet{Feiereisen1981} and \citet{Blaisdell1996}, respectively, 
in the discretization of the momentum and continuity equations to obtain 
stable simulations. The discretization of Eq.~\eqref{eq:CK}
was considered for the first time by \citet{Kennedy2008},
whereas the one in Eq.~\eqref{eq:CL} is named \textit{linear} since only 
the gradients of linear quantities appear.
Note that for the continuity equation, the forms $\mC^D$ and $\mC^{\phi}$
reduce to the classical divergence form, whereas $\mC^u$, $\mC^{\rho}$ and $\mC^L$
are equivalent to the unique advective form which can be defined for
the case of quadratic nonlinearities.
Any linear convex combination of the above mentioned forms can be equally 
considered as a consistent expression of the nonlinear convective term. 
This distinction has little importance in the continuous
formulation, since all these expressions are equivalent, once the analytical
manipulations required to derive one from the others are assumed to be valid.
However, the corresponding discretizations behave usually differently, because the
product rule, which is required to switch from one form to the others, 
does not hold in general for discrete operators.
The differences among the various forms clearly
emerge when considering the  discrete evolution  
of induced quantities, such as kinetic energy. 
As it will be recalled in the next section, a divergence structure for the nonlinear 
convective term always induces a divergence structure for the analogous
term in the evolution equation for the generalized energy $\rho\phi^2/2$.
This implies that convective terms do not contribute to the evolution of 
the total amount of generalized energy.
This property is usually lost if the discretization is not properly done,
whereas experience shows that there are beneficial effects in retaining it.
In the next section we will derive rigorous conditions under which this 
property is reproduced at discrete level when a generic linear combination 
of the forms \eqref{eq:CD}-\eqref{eq:CL} is adopted.

In what follows we will assume that the governing equations are numerically
treated by adopting a semidiscretization procedure, in which 
the equations are firstly discretized in space, and then integrated in time.
Hence, we assume that all the manipulations involving time derivatives can
be carried out at the continuous level. 
The effects of discrete time integration will not be discussed
in detail in this paper. It is worth to mention that temporal errors are typically
of dissipative character (especially when using Runge-Kutta schemes) and can be
controlled by using sufficiently small time steps \cite{capuano2017explicit}.
We will also assume that space derivatives are approximated by central schemes,
both explicit or compact, on a collocated mesh layout. 
It is well known that for this class of schemes, although the product rule is not 
valid in general, the discrete counterpart of integration by parts 
(commonly referred to as summation-by-parts, SPB) can be shown to 
hold \citep{Mansour1979}.
The general behaviour of the spatial discretization in the evolution of 
induced quantities can hence be derived by transforming the nonlinear convective terms
by only employing analytical manipulations for time derivatives and the 
SBP rule.

\section{Energy-preserving formulations}\label{Sec:Splitting}
\subsection{Derivation of the new forms}

For a generic scalar variable $\phi$, one can easily obtain the simple relation
\beq \label{eq:TimeDerRel}
\dfrac{\partial \rho\phi^2/2}{\partial t} = 
\phi\dfrac{\partial \rho\phi}{\partial t} -
\dfrac{\phi^2}{2}\dfrac{\partial \rho}{\partial t},
\eeq
whose derivation only employs manipulation of temporal derivatives.
If $\phi$ satisfies an equation of the type $\partial \rho \phi/\partial t=-\mathcal{C}$, 
the evolution equation for the generalized energy $\rho\phi^2/2$  is given by
\beq \label{eq:KinEnPhi}
\dfrac{\partial \rho\phi^2/2}{\partial t} = -\left(\phi\mC - \dfrac{\phi^2}{2}\mM\right),
\eeq
where $-\mM$ is the right-hand-side of Eq.~\eqref{eq:mass}. 
If $\mC$ has the divergence structure of Eq.~\eqref{eq:convgen}, 
Eq.~\eqref{eq:KinEnPhi} easily reduces, by analytical manipulation of the spatial derivatives,
to the equation
\beq\label{eq:KinEnPhi2}
\dfrac{\partial \rho\phi^2/2}{\partial t} = -\dfrac{\partial\rho u_i \phi^2/2}{\partial x_i}
\eeq
which shows that a divergence structure for $\mC$ and $\mM$ induces a divergence structure for
the right-hand-side of the evolution equation for the generalized energy
$\rho\phi^2/2$. This in turn implies that the \emph{global} energy
(i.e. the energy integrated over the entire domain) is always conserved
when periodic or homogeneous boundary conditions are applied.
The associated invariants are hereafter referred to as quadratic invariants.

When the evolution equation has a more complex structure, as in the cases of 
Eqs.~\eqref{eq:momentum}-\eqref{eq:entropy}, the 
considerations made above apply limited to the convective terms, which always have
a structure of the form \eqref{eq:convgen}, whereas global energy conservation is spoiled 
by viscous and pressure forces.
This property holds in the continuous case for all the balanced quantities
of Eqs.~\eqref{eq:mass}-\eqref{eq:entropy}, and its reproduction 
in the discrete equations is usually considered to be an important target
for spatial discretization.
When $\phi$ equals $u_i$ (i.e. in the case of momentum equation) the generalized 
energy is exactly the kinetic energy of the fluid, and in incompressible 
flows, for which kinetic energy is globally conserved in the inviscid limit, 
numerical methods which are able to discretely conserve kinetic energy are highly
desirable, because of their inherent nonlinear stability
\citep{Kravchenko1997,Verstappen2003,Capuano2015b}.
Also in the cases in which $\phi$ equals $E$, $e$ or $s$, 
a discretization which ensures that convective terms
do not contribute to the rate of variation of global
quantities as $\rho E^2$, $\rho e^2$ or $\rho s^2$ integrated over 
the domain is usually regarded as desirable,
and experience shows that fulfillment of these 
additional requirements typically yields additional numerical robustness.

The derivation of Eq.~\eqref{eq:KinEnPhi2} from Eq.~\eqref{eq:KinEnPhi}
employs the classical product rule for spatial derivatives, which is generally violated by 
discrete operators.
As a consequence, the divergence structure of the convective 
term in Eq.~\eqref{eq:KinEnPhi2} is in general not reproduced on a discrete level by 
the numerical approximation, and the analytically equivalent forms, 
Eqs.~\eqref{eq:CD}-\eqref{eq:CL} behave differently when discretized.
In what follows, when  a discretization reproduces the property 
that nonlinear terms do not contribute to the generalized global  energy
balance, we will term it a \emph{globally energy-preserving} discretization.
This concept is usually independent of the classical
conservative approximation property, which is related to discrete 
preservation of the linear invariants. 
Following the standard usage, we will refer to \emph{globally conservative}
discretizations for numerical approximations which are able to reproduce 
the property that the integral of the discretized convective term over 
the domain is zero. 
Local conservation is, on the other hand, achieved when the discretization 
of the convective term can be cast as difference 
of fluxes at adjacent nodes, this in turn implying global conservation through the 
telescoping property~\citep{Pirozzoli2010}.

The condition that the discretization of the nonlinear terms must satisfy not 
to spuriously contribute to the global energy balance is easily derived by 
integrating Eq.~\eqref{eq:KinEnPhi} over the entire domain, 
and by equating to zero
\beq\label{eq:CondKinCons}
\int_{\Omega} \left(\phi\mC - \frac{\phi^2}{2}\mM\right)\,\text{d}\Omega= 0.
\eeq
If one wants this condition to be satisfied by a central discretization 
for $\mC$ and $\mM$, a suitable form for $\mC$ and $\mM$ has to be chosen
among Eqs.~\eqref{eq:CD}-\eqref{eq:CL} (or among any linear combination 
of them) such that the integral in Eq.~\eqref{eq:CondKinCons} can be shown to
vanish by virtue of application of the integration by parts rule only, 
assuming that boundary terms are zero because of periodic or homogeneous 
boundary conditions.

Following the steps of \citet{Kennedy2008}, we express
the convective terms for mass and for the generic variable $\phi$ as a linear 
combination of 
different, analytically equivalent, forms \eqref{eq:CD}-\eqref{eq:CL}:
\beq
\mM = \xi\, \mM^D + \left(1-\xi\right)\mM^A, \label{eq:Mtot}
\eeq
\beq\label{eq:Ctot}
\mC = \alpha\,\mC^D + \beta\,\mC^{\phi} + \gamma\,\mC^u + \delta\,\mC^{\rho} + \varepsilon\,\mC^L,
\eeq
where $\alpha+\beta+\gamma+\delta+\varepsilon = 1$
and $\mM^D$ and $\mM^A$ are the divergence and advective forms of the nonlinear term in 
the continuity equation.
Upon substitution of Eqs.~\eqref{eq:Mtot}-\eqref{eq:Ctot} in Eq.~\eqref{eq:CondKinCons}, 
and by transforming the resulting terms by making use only
of the integration by parts rule, one can easily derive
that the fulfillment of the condition expressed by Eq.~\eqref{eq:CondKinCons} imposes the 
following constraints on the coefficients
\beq\label{eq:CoeffCond}
\left\{
\begin{array}{lcl}
\alpha      &=& 1/2-\delta \\
\beta       &=& \xi/2      \\
\gamma      &=& \delta     \\
\varepsilon &=& \dfrac{1-\xi}{2} -\delta
\end{array}
\right. .
\eeq
The system \eqref{eq:CoeffCond} defines a two-parameter family of 
discretizations having the property that
nonlinear convective terms do not contribute to the global energy
balance, as it happens for the continuous equations.
Note that by assuming that the coefficient $\xi$ is independent 
of the coefficients appearing in Eq.~\eqref{eq:Ctot}, 
we are somehow deviating from the usual assumption that the same splitting is applied 
to the continuity and to the other balance equations.
This assumption, which has been made in the past probably 
just for the sake of simplicity, is actually not 
required, and its relaxation strongly enlarges the range of possible 
energy-preserving formulations.

Split forms which are found in the literature, and for which
energy preservation has been already shown, are but two.
The first is the Feiereisen (F) form,
which is obtained by setting the free parameters $\xi = 1, \delta = 0$,
resulting in the divergence form for the continuity equation and in 
the employment of the forms $\mC^D,\mC^{\phi}$, both weighted with $1/2$, in the 
balance equation for $\phi$.
The second is the splitting obtained by uniformly weighting the 
forms $\mC^D,\mC^{\phi},\mC^u$ and $\mC^{\rho}$ with weight $1/4$. 
This last splitting, which was 
firstly considered by \citet{Kennedy2008} and 
later shown to be energy preserving by \citet{Pirozzoli2010},
is obtained by  choosing
the free parameters $\xi = 1/2, \delta = 1/4$ and will be denoted as 
KGP (Kennedy-Gruber-Pirozzoli) in the remaining part of the paper.
The Blaisdell form, which has been used in the past as an extension of the so-called
``skew symmetric'' form in the incompressible case, cannot be obtained by choosing 
specific values of $\delta$ and $\xi$, and is, in fact, not energy preserving.

The present analysis shows that the two mentioned examples are 
particular cases of a two-parameter family of energy-preserving 
forms that can be obtained by weighting the five forms of 
Eqs.~\eqref{eq:CD}-\eqref{eq:CL}. In the next section an analysis 
of this family is proposed, and new particular energy-preserving 
split formulations are introduced.

\subsection{Analysis of the new forms}\label{Sec:analysis}

Starting from the general expression given in Eq.~\eqref{eq:CoeffCond}, two special 
one-parameter families of energy-preserving discretizations can be deduced.
The first one is obtained by setting $\varepsilon=0$ in Eq.~\eqref{eq:Ctot}.
Indeed, by performing an analysis similar to that employed for the case 
of the energy preservation, it can be easily shown
that this condition is related to the possibility of attaining
a formulation which discretely preserves
the linear invariants.
In fact, the presence of the form $\mC^L$ 
in Eq.~\eqref{eq:Ctot} prevents the possibility 
of nullifying the integral of the convective term
over the entire domain by just applying the integration 
by parts rule.
In \ref{App:LocCons} it is shown that the requirement 
$\varepsilon=0$ is also a sufficient condition for 
writing the discretization in \emph{locally} conservative form
for central, explicit schemes.

When $\varepsilon = 0$, the following one-parameter
family of energy-preserving, globally conservative forms is thus obtained
\beq\label{eq:CoeffCond_epszero}
\left\{
\begin{array}{lcl}
\xi = 1-2\delta\\
\alpha = \beta = 1/2-\delta \\
\gamma = \delta\\
\varepsilon = 0
\end{array}
\right. .
\eeq
The F and KGP forms are members of this family corresponding 
to $\delta =0$ and $\delta=1/4$.
Note that this family satisfies $\xi=\alpha+\beta$, which implies
that both the $\phi$-equation and the continuity
are discretized by employing the same split form.
Hence, a necessary condition for a split form to be energy preserving and 
globally conservative of linear invariants is that the same form is employed 
for continuity and $\phi$-equation, which in turn is equivalent to require 
that $\varepsilon=0$.

Another interesting one-parameter family can be obtained by
requiring that Eq.~\eqref{eq:Ctot} has the symmetric
structure given by $\beta=\gamma=\delta$, as done by \citet{Kennedy2008}, yielding
the following one-parameter family
\beq\label{eq:CoeffCond_KG}
\left\{
\begin{array}{lcl}
\xi = 2\delta\\
\alpha = 1/2-\delta \\
\beta = \gamma =\delta\\
\varepsilon = 1/2 -2\delta
\end{array}
\right. .
\eeq
The symmetry assumption is actually not
strictly needed, since the special role played by the continuity equation 
breaks the symmetry among the  `quadratic' forms 
$\mC^{\phi},\mC^{u}$ and $\mC^{\rho}$ of the convective term. 
In this respect the family of energy-preserving 
split forms identified by \citet{Kennedy2008} does not have special significance,
but it is here highlighted to allow a comparison with the work of those authors.
In this respect, we note that in Fig.~7 of \citet{Kennedy2008} a chart 
was reported showing the output in terms of crashed or completed simulation
for test cases of compressible isotropic turbulence carried out 
with different splitting of the momentum and energy equations in the $\alpha-\beta$ plane 
(same convention for the coefficents of Eq.~\eqref{eq:Ctot} is used).
The authors' comment on that test campaign was that 
\emph{a diagonal band of $\alpha-\beta$ pairs result in the DNS code not crashing}.
That `diagonal band' infact coincides with the energy-preserving family of forms 
identified by Eq.~\eqref{eq:CoeffCond_KG}, which
is a first indirect confirmation of the validity of 
the present analysis.

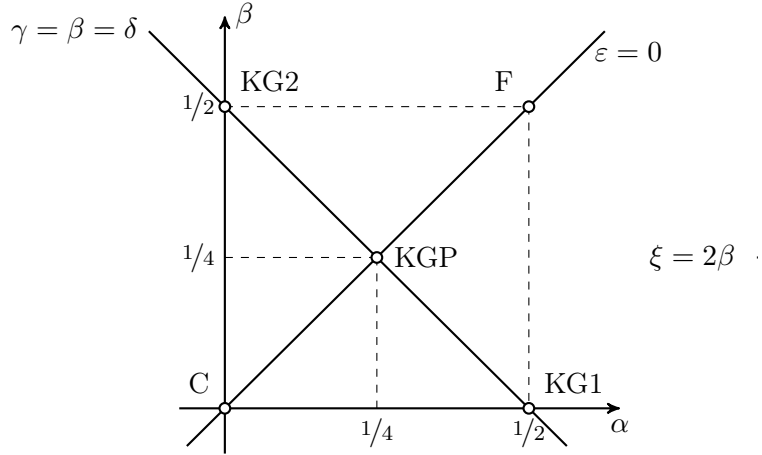
\begin{figure}
\centering
\begin{tikzpicture}[
    thick,
    >=stealth',
    dot/.style = {
      draw,
      fill = white,
      circle,
      inner sep = 0pt,
      minimum size = 4pt
    }
  ]
  \coordinate (O) at (0,0);
  \draw[->] (-0.6,0) -- (5.2,0) coordinate[label = {below:$\alpha$}] (xmax);
  \draw[->] (0,-0.6) -- (0,5.2) coordinate[label = {right:$\beta$}] (ymax);
  \path[name path=x] (0.3,0.5) -- (6.7,4.7);
  \path[name path=y] plot[smooth] coordinates {(-0.3,2) (2,1.5) (4,2.8) (6,5)};
  \draw[black]   (-0.5,-0.5) -- (5,5) node[pos=0.95, right] {$\varepsilon = 0$};
  \draw[black] (-1,5) -- (4.5,-0.5) node[pos=0.0, left] {$\gamma = \beta = \delta$};
  \draw[black,dashed,thin]       (2,2) -- (2,0) node[pos=1.0, below] {$\nicefrac{1}{4}$};
  \draw[black,dashed,thin]       (0,2) -- (2,2) node[pos=0.0, left] {$\nicefrac{1}{4}$};
  \draw[black,dashed,thin]       (4,4) -- (4,0) node[pos=1.0, below] {$\nicefrac{1}{2}$};
  \draw[black,dashed,thin]       (0,4) -- (4,4) node[pos=0.0, left] {$\nicefrac{1}{2}$};
  \draw[black]       (2,2) -- (2,2) node[dot,label = {right:KGP}] {};
  \draw[black]       (4,4) -- (4,4) node[dot,label = {above left:F}] {};
  \draw[black]       (0,0) -- (0,0) node[dot,label = {above left:C}] {};
  \draw[black]       (4,0) -- (4,0) node[dot,label = {above right:KG1}] {};
  \draw[black]       (0,4) -- (0,4) node[dot,label = {above right:KG2}] {};
  \draw             (7,2) -- (7,2) node[label = {left:$\xi = 2\beta $}] {};
\end{tikzpicture}
\caption{Chart showing two families of energy-conserving schemes. \label{fig:chart}}
\end{figure}

\begin{table}
\renewcommand\arraystretch{1.8}
\centering
\begin{tabular}{c||c|ccccc}
 &   $\xi$   & $\alpha$ &  $\beta$  & $\gamma$  & $\delta$ & $\varepsilon$ \\
\hline \hline
F   &  1  & 1/2 & 1/2 &  0  &  0  &  0  \\
C   &  0  &  0  &  0  & 1/2 & 1/2 &  0  \\
KGP & 1/2 & 1/4 & 1/4 & 1/4 & 1/4 &  0   \\
KG1 &  0  & 1/2 &  0  &  0  &  0  & 1/2  \\
KG2 &  1  &  0  & 1/2 & 1/2 & 1/2 & $-1/2$  \\
\hline \hline
\end{tabular}
\caption{Coefficients of classical and new energy-preserving split forms analyzed in Sec.~\ref{Sec:analysis}.  \label{tab:Forms}}
\end{table}

The only form belonging to both classes is the one given by substituting $\delta=1/4$
in \eqref{eq:CoeffCond_KG} or \eqref{eq:CoeffCond_epszero}, i.e. the KGP form,
which turns out to be the only globally conservative and energy-preserving form among
the ones analyzed by Kennedy and Gruber.
A chart showing the two one-parameter families
on a $\alpha-\beta$ plane is given in Fig.~\ref{fig:chart}.
In this graph the F and KGP forms are highlighted, together
with three new forms with a particularly simple structure, which are briefly outlined below.
A summary of the coefficients of these energy-preserving forms is also given in Table~\ref{tab:Forms}.
\begin{enumerate}
\item The form denoted as KG1 has parameters 
$\alpha=\varepsilon=1/2$ and $\xi=\beta=\gamma=\delta=0$.
It is energy preserving, but is not globally conservative of linear 
invariants.
The continuity equation is discretized with the advective form and the
$\mC^D$ and $\mC^L$ forms are used with weight $1/2$, 
\begin{align}
\dfrac{\partial\rho}{\partial t}&=\rho\dfrac{\partial u_j}{\partial x_j}+u_j\dfrac{\partial \rho}{\partial x_j}\\
\dfrac{\partial\rho \phi}{\partial t}&=\dfrac{1}{2}\left(\dfrac{\partial \rho u_j \phi }{\partial x_j}+
\rho \phi\dfrac{\partial u_j}{\partial x_j} + \rho u_j\dfrac{\partial \phi}{\partial x_j} +
\phi u_j\dfrac{\partial \rho}{\partial x_j}\right).\label{eq:KG1}
\end{align}
Eq.~\eqref{eq:KG1} was used by Kennedy and Gruber
in their direct numerical simulations of decaying compressible isotropic turbulence.
However, in their implementation the form corresponding to the case $\xi=1/2$ was 
adopted for continuity equation, hence the global scheme used in their paper 
is not strictly energy preserving.
\item The form denoted with KG2 has parameters 
$\xi = 1, \alpha = 0, \beta=\gamma=\delta=1/2$ and $\varepsilon = -1/2$.
As for the KG1 form, this form is energy preserving, but it does not globally preserve linear 
invariants.
The continuity equation is discretized with the divergence form and in the $\phi$-equation
$\mC^{\phi},\mC^u$ and $\mC^{\rho}$ forms are used with equal weights $1/2$, 
whereas $\mC^L$ is weighted with $-1/2$.
The $\phi$-equation has only ``quadratic'' terms, i.e. terms of the type $f\nabla gh$
\begin{align}
\dfrac{\partial\rho}{\partial t}&=\dfrac{\partial \rho u_j}{\partial x_j}\\
\dfrac{\partial\rho \phi}{\partial t}&=\dfrac{1}{2}\left(\phi\dfrac{\partial \rho u_j}{\partial x_j} +
u_j\dfrac{\partial\rho \phi}{\partial x_j}+\rho \dfrac{\partial u_j \phi}{\partial x_j}\right).\label{eq:KG2}
\end{align}
\item The form denoted as C has parameters 
$\xi = \alpha = \beta = \varepsilon = 0$ and $\gamma=\delta=1/2$.
This is an energy-preserving form which also globally preserves linear invariants.
The continuity equation is discretized with the advective form and in the momentum equation
        $\mC^{u}$ and $\mC^{\rho}$ forms are both weighted with $1/2$ 
\begin{align}
\dfrac{\partial\rho}{\partial t}&=\rho\dfrac{\partial u_j}{\partial x_j}+u_j\dfrac{\partial \rho}{\partial x_j}\\
\dfrac{\partial\rho \phi}{\partial t}&= 
\dfrac{1}{2}\left(u_j\dfrac{\partial \rho \phi}{\partial x_j} + 
\rho \phi\dfrac{\partial u_j}{\partial x_j} + 
\rho \dfrac{\partial u_j \phi}{\partial x_j} + 
\phi u_j\dfrac{\partial \rho}{\partial x_j}\right).
\end{align}
This new form is in some sense symmetric to the classical F form and seems to have 
never been considered in the literature.
Note that the one-parameter family of forms \eqref{eq:CoeffCond_epszero} can be 
equivalently expressed as a linear combination of the F and C forms with weight $\xi$ and 
$1-\xi$ respectively, the KGP form corresponding to the case $\xi=1/2$.
\end{enumerate}

An important issue related to energy-preserving 
and globally conservative discretizations
is local conservation of linear invariants.
As shown by \citet{Pirozzoli2010}, 
when central explicit finite-difference formulas of arbitrary order
are used to discretize the derivatives, both the F and the KGP 
forms can be recast in a locally conservative form, i.e. as the 
difference of numerical fluxes at successive intermediate nodes.
Besides the important implications on the convergence to weak solutions
and the improvement in computational efficiency,
this result implies that discrete local and global 
conservation of the linear invariants 
is guaranteed when the F or KGP forms are employed.
This result directly extends to the family of forms 
defined by Eq.~\eqref{eq:CoeffCond_epszero},
which may cast as a convex linear combination of F and KGP forms.
For the F and KGP forms, particularly simple and cost-effective 
flux functions have been show to exist. In~\ref{App:LocCons}
we show that similar simple flux function can also be derived for the
C form, thus estabilishing a complete analogy with the F form.
 

\section{Splitting of the energy equation}\label{Sec:EnergyEq}

Regardless of the splitting selected for the convective terms in the
continuity and momentum equations, the issue remains of which 
additional equation is most suitable among the equivalent equations 
\eqref{eq:energy}-\eqref{eq:entropy}, and which type of splitting to apply to it.
A variety of splittings of the energy equation have been considered in 
the previous literature.
\citet{Blaisdell1996} applied their splitting to the internal
energy equation, whereas \citet{Feiereisen1981}
used the evolution equation for pressure.
\citet{Kennedy2008} and \citet{Pirozzoli2010}
used total energy, although the approach adopted by Kennedy and Gruber
separately splits the convective term for $\rho E$ and the pressure
term, whereas Pirozzoli applied the splitting directly to the total enthalpy 
$E + p/\rho$.
\citet{Honein2004} applied the Feiereisen splitting to
continuity, momentum and entropy equations, and reported great advantages
in terms of robustness of the simulations.
In this section, a systematic overview of the possible approaches is presented.

In the case of vanishing viscosity, the system of Eqs.~\eqref{eq:mass}-\eqref{eq:entropy} 
can be symbolically written as
\begin{align}
\dfrac{\partial \rho}{\partial t} &= -\mathcal{M} \;, \label{eq:mass_sym} \\
\dfrac{\partial \rho u_i}{\partial t} &= -\mathcal{Q}_i -\mathcal{G}_i, \label{eq:momentum_sym} \\
\dfrac{\partial \rho E}{\partial t} &=  -\mathbb{E} -\mathcal{D}, \label{eq:energy_sym}\\
\dfrac{\partial \rho e}{\partial t} &=  -\mathcal{E} -\mathcal{P}, \label{eq:intenergy_sym}\\
\dfrac{\partial \rho s}{\partial t} &=  -\mathcal{S} , \label{eq:entropy_sym}
\end{align}
where $\mathcal{M}, \mathcal{Q}_i, \mathbb{E}, \mathcal{E}$ and $\mathcal{S}$ are the nonlinear
convective terms for $\rho,u_i,E,e$ and $s$ respectively, 
and $\mathcal{G}_i = \partial p/\partial x_i$, $\mathcal{D} = \partial pu_j/\partial x_j$, 
$\mathcal{P} = p\partial u_j/\partial x_j$.

From the definitions of $E$, $e$ and $s$ the following general relations are easily 
derived by manipulating only temporal derivatives, 
\begin{align}
\dfrac{\partial \rho E}{\partial t} &= \dfrac{\partial \rho e}{\partial t} +
\dfrac{\partial \rho u_i^2/2}{\partial t} \label{eq:RelE_e} , \\
\dfrac{\partial \rho s}{\partial t} &= \dfrac{c_v}{e} \dfrac{\partial \rho e}{\partial t} 
 + \left(s-\gamma c_v\right)\dfrac{\partial \rho}{\partial t}. \label{eq:RelE_s}
\end{align}
From these relations the effects of the spatial discretization of any quantity 
among $E, e, s$ on the balance of the other is easily obtained.
In the following sections we separately explore the possible alternatives.

\subsection{Discretization of the total energy equation}

Discretization of the continuity and momentum equations according to 
Eqs.~\eqref{eq:mass_sym}-\eqref{eq:momentum_sym} induces a discrete evolution equation for 
$\rho u_i^2/2$ of the form
\beq
\dfrac{\partial \rho u_i^2/2}{\partial t} = -\left(u_i\mQ_i-\dfrac{u_i^2}{2}\mM\right)-u_i\mG_i.
\label{eq:kinenergy_sym}
\eeq
which is analogous to Eq.~\eqref{eq:KinEnPhi} in the case $\phi = u_i$.
By construction, in the absence of pressure gradients (from now on this 
assumption will be tacitly made when referring to local and global conservation properties) kinetic energy 
is globally conserved, since energy-preserving splittings are used for $\mQ_i$ and $\mM$. 
Local conservation is however not guaranteed in general, even when
locally conservative discretizations are used for $\mQ_i$ and $\mM$.
The additional discretization of the total energy equation, Eq.~\eqref{eq:energy_sym}, 
through a locally conservative and globally energy-preserving form, 
ensures that $\rho E$ is conserved locally 
and $\rho E^2$ is conserved globally.
By virtue of Eq.~\eqref{eq:RelE_e} this in turn ensures that $\rho e$ is 
conserved globally, and it evolves through 
an equation of the form \eqref{eq:intenergy_sym}, where $\mE$ and $\mP$ 
are expressed in the form
\begin{align}
\mE &= \mbE -u_i\mQ_i+\dfrac{u_i^2}{2}\mM ,\\
\mP &= \mD - u_i\mG_i .
\end{align}
This implies that, according to Eqs.~\eqref{eq:KinEnPhi} and \eqref{eq:RelE_e},
the discrete evolution equation for $\rho e^2/2$ is 
\beq
\dfrac{\partial \rho e^2/2}{\partial t} = -e\left(\mbE-u_i\mQ_i+\dfrac{u_i^2}{2}\mM\right)+
\dfrac{e^2}{2}\mM - e\left(\mD-u_i\mG_i\right) ,
\eeq
which readily shows that $\rho e^2$ is not conserved globally
in general, i.e. a locally conservative, energy-preserving discretization
of $\rho,\rho u_i$ and $\rho E$ is globally conservative but not energy preserving for $\rho e$.

The induced equation for entropy is of the type \eqref{eq:entropy_sym}, where 
by virtue of Eq.~\eqref{eq:RelE_s}, $\mS$ has the form
\beq
\mS = \dfrac{c_v}{e} \left(\mbE -u_i\mQ_i+\dfrac{u_i^2}{2}\mM\right)
 + \left(s-\gamma c_v\right)\mM +
 \dfrac{c_v}{e} \left( \mD - u_i\mG_i\right) ,
\eeq
from which it can be readily seen that $\rho s$ is in general not conserved globally.

Note that, due to the divergence structure of the pressure term in the total energy equation,
Eq.~\eqref{eq:energy_sym} can be equivalently expressed as 
\beq\label{eq:convEnthalpy}
\dfrac{\partial \rho E}{\partial t} = -\mbE -\mathcal{D} = - \mH
\eeq
where $\mH$ has the classical structure of Eq.~\eqref{eq:convgen} with $\phi = E+p/\rho$.
We will hereafter refer to `total energy splitting' as the classical splitting of 
$\mbE$ with accompanying discretization of $\mD$ in divergence form, 
and to `total enthalpy splitting' as the splitting directly applied to $\mH$.

\subsection{Discretization of the internal energy equation}

Discretization of the internal energy equation, Eq.~\eqref{eq:intenergy_sym}, in addition
to continuity and momentum equations, through a globally (and locally) conservative and a 
globally energy-preserving discretization, of course guarantees that $\rho e$ is conserved locally, 
and $\rho e^2$ is conserved globally.
As for the previous case, by virtue of Eq.~\eqref{eq:RelE_e} this guarantees that $\rho E$ is 
conserved globally and it evolves through 
an equation of the form \eqref{eq:energy_sym} where $\mbE$ and $\mD$ 
are expressed in the form
\begin{align}
\mbE &= \mE +u_i\mQ_i-\dfrac{u_i^2}{2}\mM\label{eq:E_from_e} , \\
\mD  &= \mP + u_i\mG_i . \label{eq:D_from_P}  
\end{align}
According to Eqs.~\eqref{eq:KinEnPhi} and \eqref{eq:RelE_e},
the discrete evolution equation for $\rho E^2/2$ is 
\beq
\dfrac{\partial \rho E^2/2}{\partial t} = -E\left(\mE +u_i\mQ_i-\dfrac{u_i^2}{2}\mM\right)+
\dfrac{E^2}{2}\mM - E\left(\mP+u_i\mG_i\right), \label{eq:totenergy_sym}
\eeq
from which it is again easily seen that $\rho E^2$ is not conserved globally
in general, i.e. a locally conservative, energy-preserving discretization
of $\rho, \rho u_i$ and $\rho e$ is globally conservative but not energy preserving for $\rho E$.

The induced equation for entropy is of the type \eqref{eq:entropy_sym}, where 
by virtue of Eq.~\eqref{eq:RelE_s} $\mS$ has the form
\beq\label{eq:S_inducedby_e}
\mS = \dfrac{c_v}{e} \mE
 + \left(s-\gamma c_v\right)\mM +
 \dfrac{c_v}{e} \mP ,
\eeq
from which it can be readily seen that $\rho s$ is in general not conserved globally.

\subsection{Discretization of the entropy equation}

A discussion on the equations induced by a direct discretization of the entropy equation 
can be conducted similarly to the previous two cases. 
A discretization of Eq.~\eqref{eq:entropy_sym}, in addition
to continuity and momentum equations, through a globally (and locally) conservative and a 
globally energy-preserving discretization, guarantees that $\rho,\rho u_i$ and $\rho s$ are 
conserved locally, and $\rho u_i^2$ and $\rho s^2$ are conserved globally.
By virtue of Eq.~\eqref{eq:RelE_s} and of Eqs.~\eqref{eq:E_from_e} and \eqref{eq:D_from_P},
this  implies that $\rho e$ and $\rho E$ 
evolve through equations of the form \eqref{eq:intenergy_sym} and \eqref{eq:energy_sym}
respectively, where $\mE+\mP$ and $\mbE+\mD$
are expressed in the form
\begin{align}
\mE  + \mP &= \dfrac{e}{c_v} \mS-\dfrac{e}{c_v}\left(s-\gamma c_v\right)\mM\label{eq:e_inducedby_s} , \\
\mbE + \mD &= \dfrac{e}{c_v} \mS-\dfrac{e}{c_v}\left(s-\gamma c_v\right)\mM\label{eq:E_inducedby_s} 
+ u_i\mQ_i -\dfrac{u_i^2}{2}\mM+u_i\mG_i,
\end{align}
from which it is seen that in general, neither $\rho e$ nor $\rho E$ are globally conserved.
Note that the application of Feiereisen splitting to Eqs.~\eqref{eq:e_inducedby_s} and 
\eqref{eq:E_inducedby_s} yields exactly the non-viscous versions of Eqs.~(18) and (19) of 
\citet{Honein2004}.

\begin{table}
\renewcommand\arraystretch{1.8}
\centering
\begin{tabular}{cc|ccccccccc}
&&\multicolumn{9}{c}{Conserved variable}\\
&   &   $\rho$   & $\rho u_i$ &  $\rho E$  & $\rho e$  & $\rho s$ & $\rho u_i^2$ & $\rho E^2$ & $\rho e^2$ & $\rho s^2$\\
\hline
\multirow{4}{2cm}{Discretized \\ energy \\ equation} & $\rho E$ & $\bigodot$ & $\bigodot$ & $\bigodot$ & $\bigcirc$& $\times$ & $\bigcirc$   & $\bigcirc$ & $\times$   & $\times$  \\
&$\rho e$ & $\bigodot$ & $\bigodot$ & $\bigcirc$ & $\bigodot$& $\times$ & $\bigcirc$   & $\times$ & $\bigcirc$   & $\times$  \\
&$\rho s$ & $\bigodot$ & $\bigodot$ & $\times$   & $\times$  & $\bigodot$ & $\bigcirc$   & $\times$ & $\times$   & $\bigcirc$  \\
&$\rho e$ (dyn) & $\bigodot$ & $\bigodot$ & $\bigcirc$   & $\bigodot$  & $\bigcirc$ & $\bigcirc$   & $\times$ & $\bigcirc$   & $\times$  \\
&$\rho s$ (dyn) & $\bigodot$ & $\bigodot$ & $\bigcirc$   & $\bigcirc$  & $\bigodot$ & $\bigcirc$   & $\times$ & $\times$   & $\bigcirc$  \\
\hline
\end{tabular}
\caption{Conservation properties induced by different energy balance equations discretized in split form. 
$\bigodot$: variable conserved locally and globally,
$\bigcirc$: variable conserved globally but not locally, 
$\times$: variable not conserved. \label{tab:ConvProp}}
\end{table}

\subsection{Adaptive selection of the split form \label{sec:adaptive}}

The results of the above made considerations are summarized in Table~\ref{tab:ConvProp},
showing that by discretizing directly the internal or the total energy equation,
global conservation of entropy is not guaranteed. On the other hand, discretizing the 
entropy equation guarantees that $\rho s$ is conserved locally and $\rho s^2$ globally,
but conservation of the total and internal energy is lost, even in global sense.

Since the possible split forms which are energy preserving and globally (and locally) conservative
of linear invariants constitute a one-parameter family, it is tempting to exploit
the degree of freedom given by the free parameter in order to satisfy additional conservation
properties.
As shown below, this can be achieved under certain conditions, through an adaptive procedure
that selects the splitting within the family in a dynamic way, by enforcing 
an additional global conservation constraint.
This procedure can in fact be designed in different ways. Here we provide some possibilities
to illustrate the general idea.

Let us consider the case in which the continuity and momentum equations are discretized 
together with the internal energy equation with a globally (and locally) conservative 
discretization and with an energy-preserving split form.
According to Eq.~\eqref{eq:S_inducedby_e},
the condition for global conservation of entropy is given by
\beq\label{eq:Conserv_S}
\int_{\Omega} \left(\dfrac{c_v}{e} \mE
 + \left(s-\gamma c_v\right)\mM +
 \dfrac{c_v}{e} \mP\right)\,\text{d}\Omega=0 .
\eeq
Since $\rho$ is globally conserved, the space integral of $\mM$ is zero,
and Eq.~\eqref{eq:Conserv_S} reduces to
\beq\label{eq:Conserv_S_2}
\int_{\Omega} \left(\dfrac{c_v}{e} \mE
 + s\mM +
 \dfrac{c_v}{e} \mP\right)\,\text{d}\Omega=0 .
\eeq
Note that, if $\mM$ and $\mE$ are discretized through a split form of the family 
\eqref{eq:CoeffCond_epszero}, they may be expressed as
\begin{align}
\mM &= \xi\mM^D + (1-\xi)\mM^A \label{eq:Msplit}, \\
\mE &= \xi\mE^F + (1-\xi)\mE^C \label{eq:Esplit},
\end{align}
where $\mE^F$ and $\mE^C$ are the convective terms of the internal energy equation 
discretized in the F and in the C form, respectively.
By substituting Eqs.~\eqref{eq:Msplit} and \eqref{eq:Esplit} into 
Eq.~\eqref{eq:Conserv_S_2} one is left with
\beq\label{eq:Dynamic1}
\int_{\Omega}\xi \left[\dfrac{c_v}{e}\left(\mE^F-\mE^C\right)+s\left(\mM^D-\mM^A\right) \right] + 
\left(\dfrac{c_v}{e}\mE^C+s\mM^A+\dfrac{c_v}{e}\mP\right)\,\text{d}\Omega=0,
\eeq
From Eq.~\eqref{eq:Dynamic1} the free parameter $\xi$ can be selected in order to satisfy 
Eq.~\eqref{eq:Conserv_S_2}.
In fact, dynamically adjusting in time $\xi$ according to
\beq\label{eq:Dynamic1_xi}
\xi_e = -\dfrac{\int_{\Omega}\left(\dfrac{c_v}{e}\mE^C+s\mM^A+\dfrac{c_v}{e}\mP\right)\,\text{d}\Omega } 
{\int_{\Omega}\dfrac{c_v}{e}\left(\mE^F-\mE^C\right)+s\left(\mM^D-\mM^A\right)\,\text{d}\Omega}
\eeq
guarantees that Eq.~\eqref{eq:Conserv_S_2} is satisfied at each time instant, and the procedure will
conserve locally $\rho, \rho u_i$ and $\rho e$ and globally $\rho u_i^2,\rho E,\rho e^2$ and $\rho s$.

Similarly, if one considers the case in which the entropy equation is discretized together 
with the continuity and the momentum equations, according to Eq.~\eqref{eq:e_inducedby_s},
the condition for global conservation of internal is given by
\beq\label{eq:Conserv_e}
\int_{\Omega} \left(\dfrac{e}{c_v} \mS
 - \dfrac{e}{c_v} \left(s-\gamma c_v\right)\mM - \mP\right)\,\text{d}\Omega=0 .
\eeq
By expressing $\mM$ through Eq.~\eqref{eq:Msplit} and $\mS$ as $\mS=\xi\mS^F+(1-\xi)\mS^C$, Eq.~\eqref{eq:Conserv_e} reduces to 
 \beq\label{eq:Dynamic2}
 \int_{\Omega} \xi\dfrac{e}{c_v}\left[\left(\mS^F-\mS^C\right) - \left(s-\gamma c_v\right)\left(\mM^D-\mM^A\right)\right]
  + \dfrac{e}{c_v}\left(\mS^C-\left(s-\gamma c_v\right)\mM^A\right) -  \mP\,\text{d}\Omega=0 ,
 \eeq
from which one may infer that dynamically adjusting in time $\xi$ according to
\beq\label{eq:Dynamic2_xi}
\xi_s = -\dfrac{\int_{\Omega}\left(\dfrac{e}{c_v}\left(\mS^C-\left(s-\gamma c_v\right)\mM^A\right) -  \mP\right)\,\text{d}\Omega } 
{\int_{\Omega} \dfrac{e}{c_v}\left[\left(\mS^F-\mS^C\right) - \left(s-\gamma c_v\right)\left(\mM^D-\mM^A\right)\right]  \,\text{d}\Omega}
\eeq
guarantees that Eq.~\eqref{eq:Conserv_e} is satisfied at each time instant, and the procedure will
conserve locally $\rho, \rho u_i$ and $\rho s$ and globally $\rho u_i^2,\rho s^2, \rho e$ and $\rho E$.

\section{Numerical tests: the inviscid Taylor-Green flow}\label{Sec:Tests}

In this section, the inviscid compressible Taylor-Green flow is 
used as a test case for comparing the performance of the 
various split forms analyzed in Sec.~\ref{Sec:analysis}. 
Different splittings of the energy equation are also considered
as explained in Sec.~\ref{Sec:EnergyEq}, 
resulting in a test matrix of $20$ different formulations.
We should point out that for convenience of computational implementation,
and as suggested by \citet{Honein2004} the total energy equation is solved in all cases,
however with right-hand side rearranged either according to Eqs.~\eqref{eq:E_from_e}, \eqref{eq:D_from_P}
to emulate splitting of the internal energy equation, or to Eq.~\eqref{eq:E_inducedby_s},
to emulate splitting of the entropy equation.
Spatial discretization is performed in all cases by standard explicit central schemes of 
order $2$, $4$ and $6$, and time integration is 
carried out by means of the third-order TVD Runge-Kutta scheme of 
\citet{ShuOsher1989} and by the standard fourth-order Runge-Kutta scheme (RK4).
The flow is integrated in a triply periodic cube of size $2\pi$ with 
zero viscosity, with the following initial conditions~\citep{JohnsenJCP2010}
\begin{align*}
    \rho &= 1 , \\
    u    &= \sin(x)\cos(y)\cos(z)  , \\
    v    &= -\cos(x)\sin(y)\cos(z)  , \\
    w    &= 0  , \\
    p    &= 100+\dfrac{\left(\cos(2x)+\cos(2y)\right)\left(\cos(2z)+2\right)-2}{16} ,
\end{align*}
where pressure is taken sufficiently high to provide a flow which is effectively
incompressible. The ratio of specific heats $\gamma$ is set to $1.4$.
It is well known that, after an initial transient the initially smooth flow
experiences distortion and stretching, and it quickly undergoes instabilities 
characterized by the formation of smaller and smaller scales.
For any given grid, after a sufficiently long time interval the flow develops 
unresolved scales, entering a thermalized random regime.
This behaviour makes this flow a convenient testbed to check the stability of numerical methods 
in strongly under-resolved situations, and the absence of viscosity allows rigorous
verification of the conservation of the invariants of motion.

\begin{table}
    \centering
    \begin{tabular}{|c|c|c|c|c|c|c|c|}
      \toprule
                              &  &   Order    &  \multicolumn{5}{c}{Forms for convective terms} \\
        \cmidrule{2-8} 
                              &  &            &     KGP    &      F      &      C     &    KG1     &     KG2    \\
        \cmidrule{2-8} 
\multirow{12}{2cm}{Formulation \\ of energy \\ equation} 
        &      \multirow{3}{*}{Internal energy} & 2  & \checkmark &  $\times$   &  $\times$  &  $\times$  &  $\times$  \\
        &                        & 4  & \checkmark &  $\times$   &  $\times$  &  $\times$  &  $\times$  \\
        &                        & 6  & \checkmark &  $\times$   &  $\times$  &  $\times$  &  $\times$  \\
        \cmidrule{2-8} 
        &      \multirow{3}{*}{\shortstack{Total energy with \\ total energy splitting }}& 2  & \checkmark &  $\times$   &  $\times$  &  $\times$  &  $\times$  \\
        &                        & 4  & \checkmark &  $\times$   &  $\times$  &  $\times$  &  $\times$  \\
        &                        & 6  & \checkmark &  $\times$   &  $\times$  &  $\times$  &  $\times$  \\
        \cmidrule{2-8} 
        &      \multirow{3}{*}{\shortstack{Total energy with \\ total enthalpy splitting}}& 2  & \checkmark & \checkmark  & \checkmark &  $\times$  &  $\times$  \\
        &                        & 4  & \checkmark & \checkmark  & \checkmark &  $\times$  &  $\times$  \\
        &                        & 6  & \checkmark & \checkmark  & \checkmark &  $\times$  &  $\times$  \\
        \cmidrule{2-8} 
        &      \multirow{3}{*}{Entropy} & 2  & \checkmark & \checkmark  & \checkmark &  $\times$  &  $\times$  \\
        &                        & 4  & \checkmark & \checkmark  & \checkmark &  $\times$  &  $\times$  \\
        &                        & 6  & \checkmark & \checkmark  & \checkmark &  $\times$  &  $\times$  \\
        \bottomrule
    \end{tabular}
    \caption{Test matrix for inviscid Taylor-Green flow numerical simulations. \checkmark numerically stable; $\times$ numerically unstable. Refer to Table~\ref{tab:Forms} for the definition of the various split forms.} \label{tab:convergence}
\end{table}

\subsection{Robustness assessment}\label{Sec:Robustness}

A numerical simulation campaign was first carried out to verify the computational robustness 
of the various formulations.
The Euler equations are discretized on a $32^3$ uniform grid
and integrated in time with the third-order TVD Runge-Kutta scheme
up to the final time $t = 256$ with $\text{CFL}=1$.

The results of the tests are summarized in Table~\ref{tab:convergence},
in terms of numerical stability or instability, at least within the time integration interval.
The data have been checked to be sufficiently general by performing additional spot calculations 
at different CFL numbers and changing the time integration scheme to RK4.
The table shows that among the various split forms, the KGP form is most robust,
allowing to achieve stable computations when applied in conjunction with any of the formulations 
for the energy equation.
Furthermore, use of the total energy equation with total enthalpy splitting or of the entropy equation
is found to be the most robust choice for the energy equation, yielding numerical stability 
also for the F and C split forms.
In all other cases the simulations diverged within the time integration interval.
The order of accuracy of the spatial discretization seems to have no influence 
on stability, although it is clearly expected to play a role on the accuracy of the solution.

\begin{figure}
    \centering
    \includegraphics[width=1\linewidth]{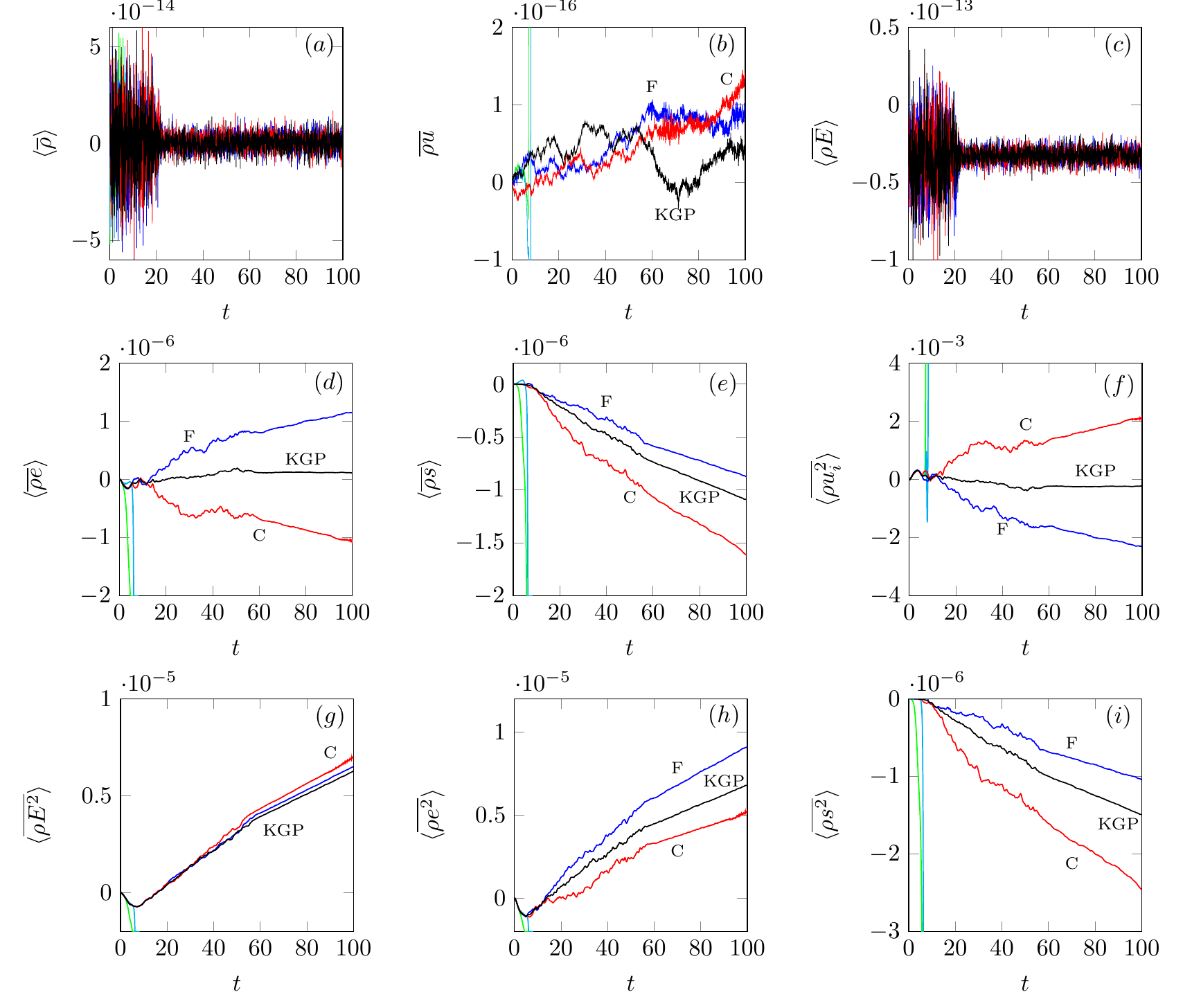}
    \captionof{figure}{Time evolution of linear and quadratic invariants for inviscid Taylor-Green flow 
    using different splittings of the convective terms, with second-order central discretization of 
    the space derivatives.
    The total energy equation is used with total enthalpy splitting.}
    \label{fig:2}
\end{figure}

\begin{figure}
    \centering
    \includegraphics[width=1\linewidth]{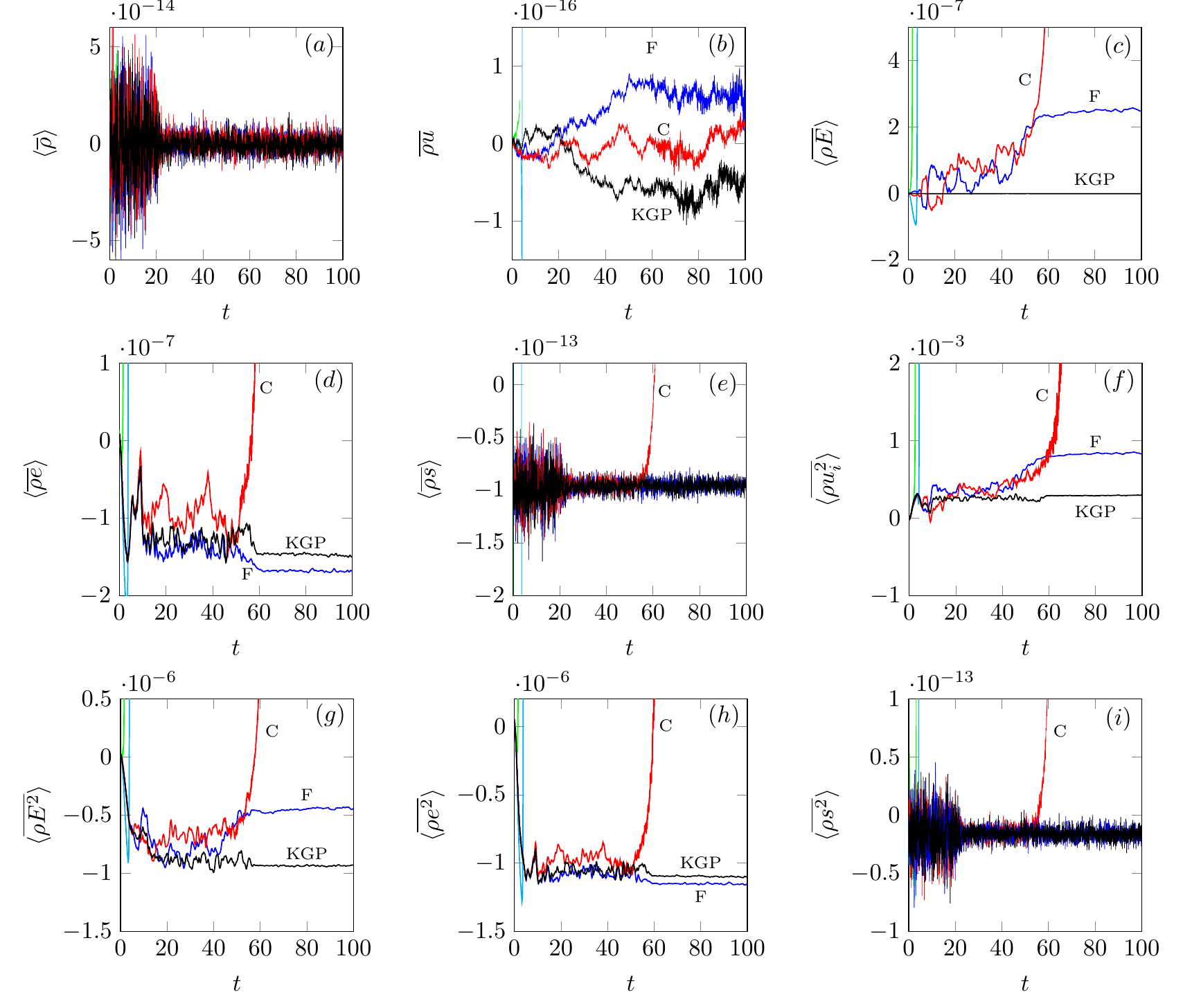}
    \captionof{figure}{Time evolution of linear and quadratic invariants for inviscid Taylor-Green flow 
    using different splittings of the convective terms, with second-order central discretization of 
    the space derivatives.
    The entropy energy equation is used.}
    \label{fig:3}
\end{figure}

In order to verify the predictions developed in the previous sections, 
selected calculations have been carried out with the more accurate RK4 
scheme and with $\text{CFL}=0.1$, in order to reduce temporal errors as much as possible.
The results are shown in Fig.~\ref{fig:2}, \ref{fig:3} in terms of the
time evolution of linear and quadratic invariants.
In these plots, the overbar denotes spatial integration over the entire domain, while the brackets 
indicate normalization of the deviation from the initial value over the initial value itself
(i.e. $\langle\overline{f}\rangle = (\overline{f}-\overline{f_0})/\overline{f_0}$).

The data reported in Fig.~\ref{fig:2} are obtained by integrating the equations of mass 
momentum and total energy with total enthalpy splitting, by testing all the energy-preserving 
split forms discussed in Sec.~\ref{Sec:Splitting}. 
Central second-order approximations are used for all the space derivatives.
We find that the formulations employing the KG1 and KG2 forms diverge before $t=10$, 
and the corresponding curves are seen as vertical lines (not labelled). 
On the other hand, the energy-preserving and locally-conservative 
forms KGP, F and C are found to be stable over the entire integration interval.
As seen in panels $(a)-(c)$, 
global values of  mass, momentum and total energy are accurately conserved in 
time, up to machine precision (all calculations have been carried out using double-precision arithmetics).
Although globally preserved by convection, $\rho e$ and $\rho u_i^2$
are not conserved in time as their evolution is also affected by exchange of energy
through the exchange terms $\mP$, $u_i \mG_i$ 
(see Eqs.~\eqref{eq:intenergy_sym}, \eqref{eq:kinenergy_sym}, respectively).
Similarly, the quantity $\overline{\rho E^2}$, although preserved by convective terms, is
affected by the pressure-type term $E\left(\mP+u_i\mG_i\right)$, 
which causes slight increase in time. 
On the other hand, $\overline{\rho s}$ and $\overline{\rho s^2}$,
should stay constant in the inviscid case, hence
their variations are entirely attributable to lack of global discrete conservation.

The data reported in Fig.~\ref{fig:3} support this analysis.
In this case, splitting is applied to the mass, momentum and entropy 
equations, again with all the energy-preserving forms discussed in Sec.~\ref{Sec:Splitting}. 
The plots clearly show that mass, momentum and entropy are exactly globally conserved in time, 
together with $\overline{\rho s^2}$. 
The total energy displays 
slight deviations from constancy, because the entropy formulation does not guarantee global preservation
of total and internal energy. 
Note that for this formulation of the energy equation the C split form 
performs worse than the KGP and the F forms,
as larger deviations from the expected behaviour are observed
starting at $t\simeq 50$; however, no divergence is observed.
At $t = 100$, $\langle\overline{\rho u^2}\rangle, \langle\overline{\rho E}\rangle$ 
and $\langle\overline{\rho s}\rangle$
reach values of $0.22$, $3.8\times 10^{-4}$ and $4.1\times10^{-10}$, respectively.
Note also that, in contrast to what observed in Fig.~\ref{fig:2},
the deterioration of the performances of the C form 
also affects the quantities $\overline{\rho s}$ and $\overline{\rho s^2}$, 
which should be conserved by construction
(see the C curves in panels (e) and (i)).
This behaviour may be traced to the previously noted use of 
a surrogate total energy equation in the place of the entropy equation.
In this formulation the entropy is strictly a 
derived variable (it is evaluated from $\rho E$ through 
$\rho s = \rho c_v\ln\left(\left(\gamma-1\right)\left(\rho E-\rho u_i^2/2\right)/\rho^{\gamma -1}\right)$,
and therefore, accumulation of numerical errors may also spoil
variables which should be globally conserved by construction.
The deterioration of the accuracy of the C form, on the other side, does not affect 
the global conservation of mass and momentum, whose equations are directly solved for
(see panels (a) and (b)).

A further observation on Fig.~\ref{fig:3} relates to
the total energy evolution obtained from the KGP split form applied to the entropy 
formulation. The curve labelled as KGP in Fig.~\ref{fig:3}(c) actually shows that, in contrast to the
C and F curves, in the entropy formulation the total energy is globally conserved with good 
accuracy during the whole integration interval (its maximum absolute value is 
about $2\times 10^{-10}$). This additional conservation property of the KGP split
form in the context of the discretization of the entropy equation,
is further investigated in the forthcoming section.

\subsection{Test of adaptive splitting procedures}

With the aim of further analyzing the properties of the various split forms 
in connection to the enforcement of additional balance equations, 
the adaptive procedures proposed in Sec.~\ref{sec:adaptive}
have been tested.
The test case and the space and time discretization setup are the same 
as in Sec.~\ref{Sec:Robustness}, but the split 
form is now dynamically adjusted within the family of Eq.~\eqref{eq:CoeffCond_epszero},
by adapting the value of $\xi$ according to either
Eq.~\eqref{eq:Dynamic1_xi} or Eq.~\eqref{eq:Dynamic2_xi}.

\begin{figure}
    \centering
    \includegraphics[width=0.9\linewidth]{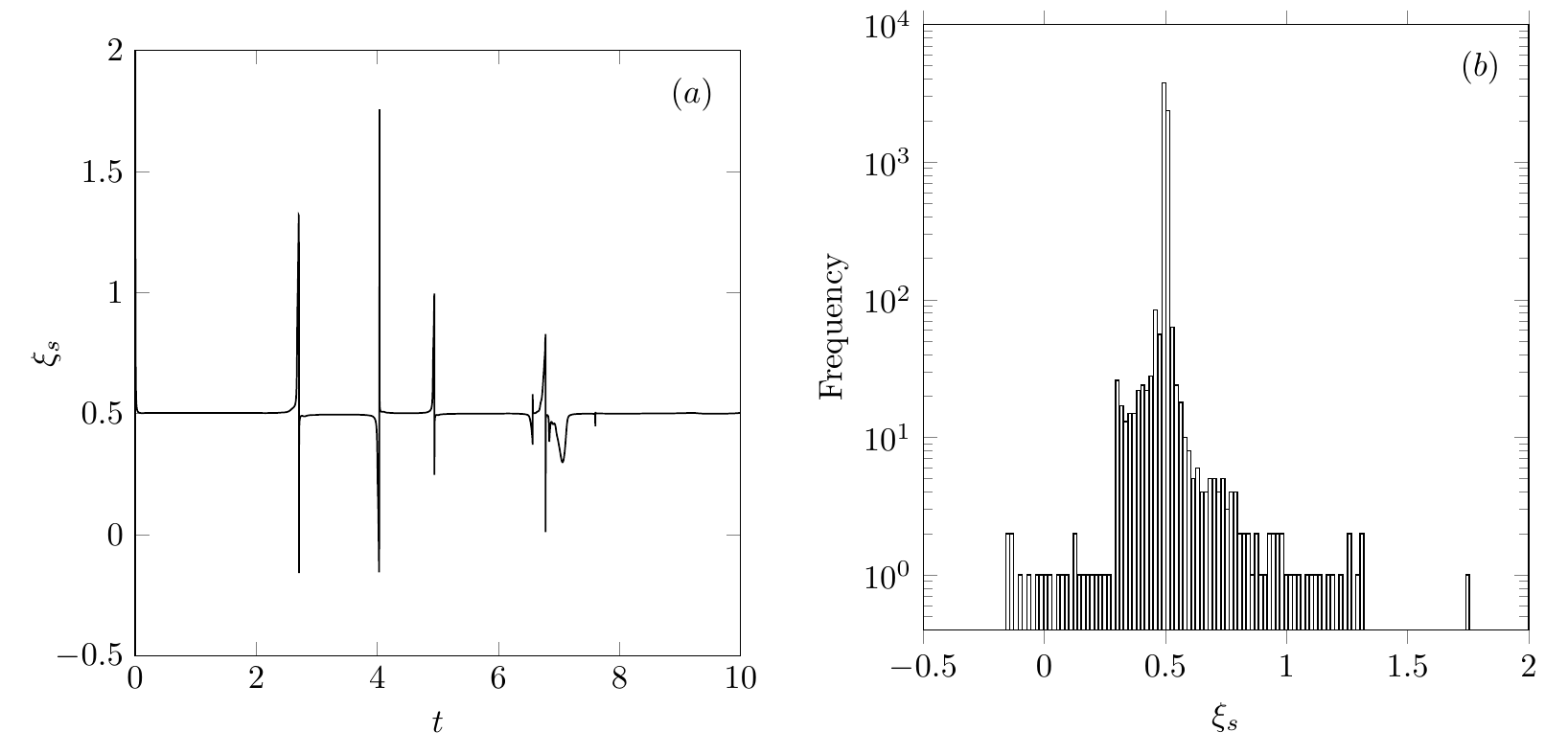}
    \captionof{figure}{Time evolution (a) and frequency distribution histogram (b) of the 
    $\xi_s$ coefficient for the adaptive procedure of Eq.~\eqref{eq:Dynamic2_xi}.}
    \label{fig:4}
\end{figure}
In Fig.~\ref{fig:4}(a)
the value of the coefficient $\xi_s$ obtained from Eq.~\eqref{eq:Dynamic2_xi} is shown
as a function of time.
In this simulation the mass, momentum and entropy equations are solved 
in split form dynamically adjusted to guarantee the additional global preservation of $\rho e$.
From this plot it is seen that, with the exception of isolated spikes
due to the possibly singular
character of Eq.~\eqref{eq:Dynamic2_xi},
the value of $\xi_s$ obtained from the dynamic procedure settles around $0.5$.
In practice, this stands to indicate that the KGP form applied to the entropy equation 
also guarantees global conservation of the internal energy, and as a consequence
global conservation of $\rho E$.
This finding is in perfect agreement with the results obtained in the previous 
section with the KGP form applied to the entropy equation (recalling Fig.~\ref{fig:3}(c)),
which was shown to yield negligible variation of global total energy.
The frequency distribution of $\xi_s$ is shown in Fig.~\ref{fig:4}(b),
where bars outside the interval $\left[-0.5,2\right]$ are not displayed.
From these data it may be estimated that $\xi_s$ falls in the         
interval $0.5\pm 0.01$ in the $90\%$ of the cases, and in the 
interval $0.5\pm 0.05$ in more than $95\%$ of the cases 
(note that semi-logarithmic representation is used).

\begin{figure}
    \centering
    \includegraphics[width=0.9\linewidth]{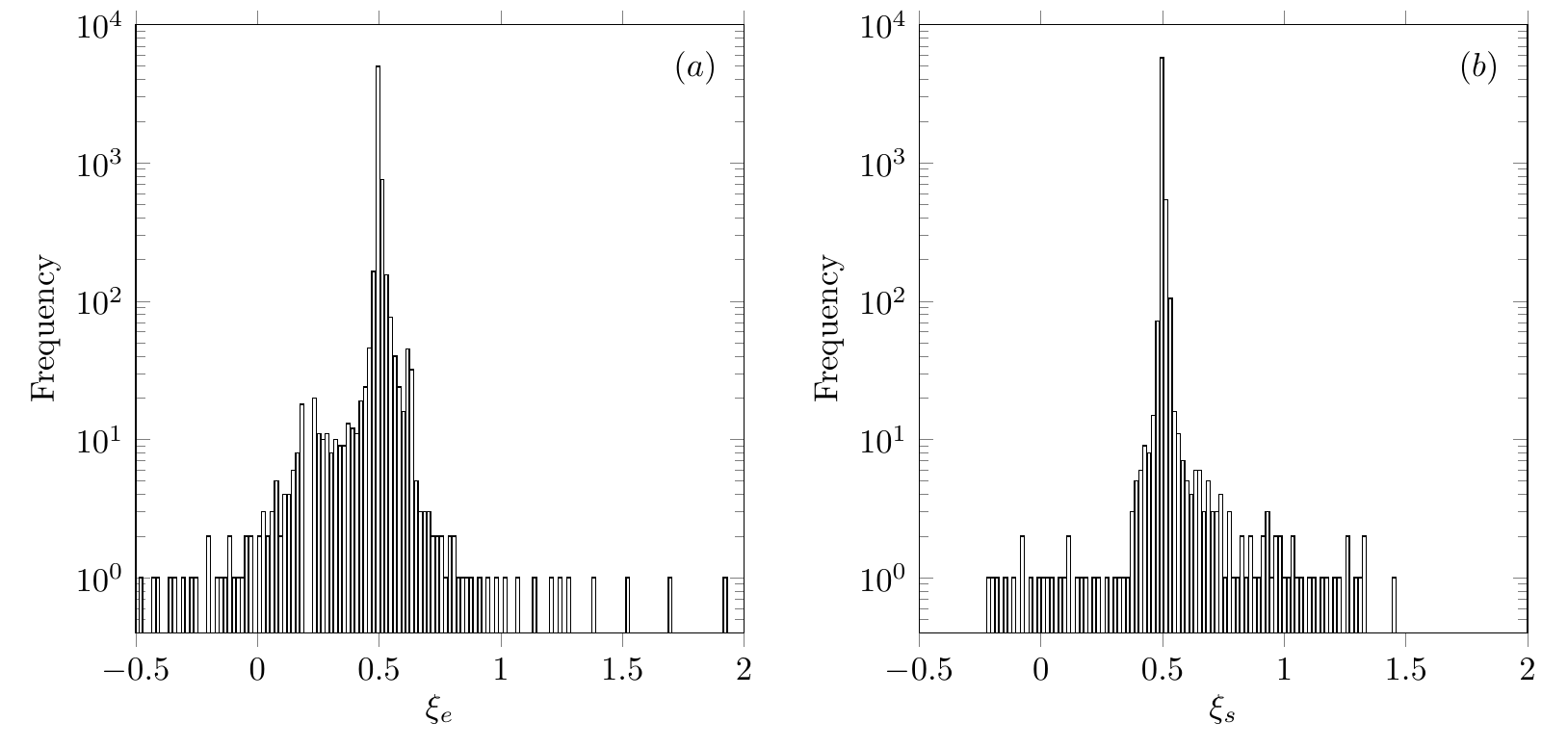}
    \captionof{figure}{Frequency distribution histograms of the 
    $\xi_e$ coefficient for the adaptive procedure of Eq.~\eqref{eq:Dynamic1_xi}
    for inviscid Taylor-Green flow (a) and 
    of $\xi_s$ for the adaptive procedure of Eq.~\eqref{eq:Dynamic2_xi}
    for viscous Taylor-Green flow at Re=1600 (b).}
    \label{fig:5}
\end{figure}

A similar situation also occurs for the dynamical procedure applied to 
the internal energy equation.
In this case the mass, momentum and internal energy equations 
are integrated in time, with a split form dynamically 
selected by the value of $\xi_e$ given by Eq.~\eqref{eq:Dynamic1_xi},
which guarantees additional global preservation of $\rho s$.
Figure~\ref{fig:5}(a) shows the frequency distribution histogram of $\xi_e$.
The convergence to the $0.5$ value corresponding to the KGP form 
is confirmed also for this procedure,
and cases in which $\xi_e$
falls in the interval $0.5\pm 0.05$ are estimated to be around $92\%$ of the total.

As far as additional global conservation properties are concerned, 
both simulations show that the KGP form is 
almost optimal over the entire time integration interval.
This includes smooth states in the initial transient, as well as
fully thermalized states in later stages.
The robustness of this finding has been further investigated
by applying the dynamic procedure to a viscous calculation. 
The same initial condition and the same domain size are employed 
for numerical integration of the viscous Taylor-Green flow on 
a $32^3$ grid at Reynolds number of $1600$.
The mass, momentum and entropy equations are solved, and the split form is adaptively
determined through application of Eq.~\eqref{eq:Dynamic2_xi}.
In Fig.~\ref{fig:5}(b) the frequency distribution histogram 
of the dynamically calculated coefficient $\xi_s$ is reported.
The histogram again shows strong tendency of the dynamical
procedure to select values of $\xi_s$ around $0.5$.
In this case the number of occurrences of $\xi_s$ 
in the interval $0.5\pm 0.05$  is around $97\%$
of the total.

\subsection{Effect of formulations of the energy equation}

The effect of the formulation used for the energy equation on the reliability 
of numerical simulations remains to be explored. 
Indeed, we find the KGP split form
to be equally robust, regardless of the energy formulation employed 
(among those introduced in Sec.~\ref{Sec:EnergyEq}), whereas the F and C forms 
proved to be stable only when either the total energy equation 
(with total enthalpy splitting) or the entropy equation 
are used.
In this section we attempt to give additional insights regarding the accuracy 
of the various (stable) formulations, with the aim of establishing the most reliable.

\begin{figure}
    \centering
    \includegraphics[width=0.9\linewidth]{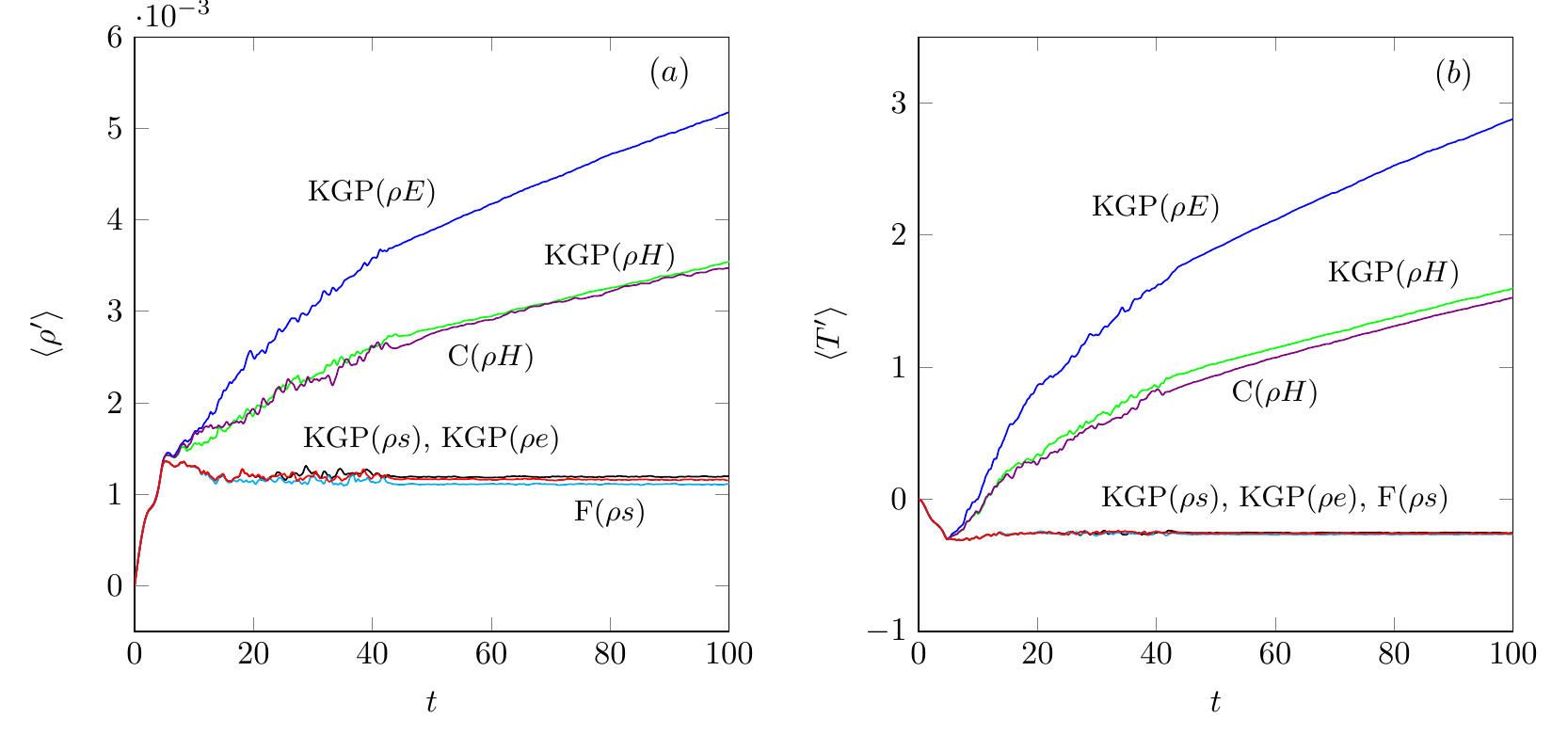}
    \captionof{figure}{Density (a) and temperature (b) fluctuations for inviscid Taylor-Green flow
    using different formulations of the energy equation coupled with different splittings.
    }
    \label{fig:6}
\end{figure}

As a measure for reliability, we monitor
the evolution of thermodynamic fluctuations in time, and in particular we consider
density and temperature fluctuations.
It is expected that after an initial transient,
these quantities level off to a constant value, 
similarly to what reported for inviscid isotropic homogeneous turbulence~\citep{Honein2004,Pirozzoli2010}. 
In Fig.~\ref{fig:6}, the r.m.s. density and temperature fluctuations 
obtained with selected numerical simulations are shown.
The parameters of the simulations are the same of Fig.~\ref{fig:2} 
and \ref{fig:3}, except for the space discretization, which is now 
carried out through fourth-order central explicit formulas.
All formulations for the energy equation are considered along with the KGP splitting
(the corresponding cases are denoted as KPG(X)),
whereas the total energy equation with total energy splitting is used with the C splitting 
(the corresponding case is denoted as C($\rho H$)),
and the entropy equation is used with the F splitting
(the corresponding case is denoted as F($\rho s$)).
Note that the two formulations KGP($\rho H$) and F($\rho s$)
match those considered by \citet{Pirozzoli2010}
and by Honein \& Moin \cite{Honein2004}, respectively,
although in the latter case compact differencing was used for 
space discretization.

The data in Fig.~\ref{fig:6} show that not all stable formulations 
yield asymptotic stabilization.
Numerical simulations with the KGP form using the total energy equation 
(irrespective of the use of total energy or total enthalpy splitting)
exhibit non-negligible increase of density and temperature fluctuations,
which, however, does not lead to numerical divergence.
Similar results are obtained with the F and C split forms in conjunction 
with the total energy equation, although only the C($\rho H$) formulation is displayed. 
This behaviour was previously highlighted by \citet{Pirozzoli2010}.
On the other hand, application of the KGP or F split forms together with the entropy 
equation provides much better results, both formulations
yielding an asymptotically constant level of the fluctuations. 
Quite surprisingly, the KGP splitting applied to the
internal energy equation also yields good results, 
as it mantains the thermodynamic fluctuations  
to a constant value close to the one given by other formulations.
This result can be justified by observing again that the KGP($\rho e$)
formulation guarantees the additional conservation of $\overline{\rho s}$.

\section{Conclusions}

A general framework for the derivation of energy-preserving split forms
for convective terms in the compressible Navier-Stokes 
equations has been presented.
In contrast to the incompressible case, for which the skew-symmetry
of the discretized convective operator was shown to be the essential 
ingredient for global preservation of kinetic energy,
energy-preserving formulations in the compressible case were not 
completely characterized.
The theory herein developed fills this gap and provides a wide 
generalization of existing split forms for the compressible 
Navier-Stokes equations.
The classical \citet{Feiereisen1981} splitting, 
and the more recent form introduded by 
\citet{Kennedy2008} and employed by \citet{Pirozzoli2010} 
(here referred to as KGP form), have been shown to be just two particular cases of 
a two-parameter family of energy-preserving forms.
The analysis of the new forms has been conducted by considering 
also the topic of global conservation of primary quantities.
It has been shown that even in the case in which the preservation of 
both linear and quadratic invariants is required, the set of admissible 
split forms constitutes a one-parameter family.
Locally conservative formulations, on the other hand, 
have shown to be possible for this restricted class (for central explicit
schemes of arbitrary order) by extending the approach used by \citet{Pirozzoli2010},
and particularly simple and economic flux functions have been derived 
also for the new splitting forms.

In compressible flows the choice of a suitable energy-preserving 
split form for the discretization of convective terms does not strictly guarantee
nonlinear stability. Another important and influential topic is the 
choice of which energy equation is most suitable among the 
various equivalent possibilities.
A systematic analysis of the induced conservation properties 
of each formulation has been presented, and the employment of the free parameters
stemming from the proposed theory on the split forms has led to the idea of `dynamic' 
procedures which are able to provide additional discrete conservation properties 
by adaptively selecting the split form within a one-parameter family.

Numerical tests on the inviscid compressible Taylor Green flow 
confirmed the theoretical predictions and provided 
new insights toward the selection of an optimal 
formulation in terms of stability and reliability.
The numerical experiments showed that global conservation of linear invariants 
is a important issue, since energy-preserving formulations which 
are not globally conservative of primary variables are typically
unstable in the nonviscous case. This result is somehow in contrast to what 
happens in incompressible flows, for which kinetic energy conservation alone
is typically  sufficient to prevent instabilities 
arising from the accumulation of aliasing errors.
Among the various split forms, the newly derived C form
has robustness properties analogous to the classical F form, while
the KGP form has shown to be the most robust,
in conjunction with all the energy equation formulations. 
This behaviour has been confirmed by 
the applicaton of adaptive procedures, which revealed that 
the KGP form is the optimal choice also with respect to additional
induced conservation requirements. 
Finally, an analysis conducted on thermodynamic fluctuations
has confirmed that the conservation of global entropy is an important ingredient.
This is true not only for robusteness, but also with respect to reliability of 
simulations, since the unphysical increase in the amplitude of the fluctuations, 
arising in some stable formulations, is not present when global entropy is 
preserved, both as a primary effect of the direct discretization of the entropy equation
or as an hidden advantage of the adopted splitting.

\appendix
\section{Locally conservative formulations}\label{App:LocCons}

By assuming a central finite difference explicit differentiation formula of the form
$$\left.\widehat{\dfrac{\partial \phi}{\partial x}}\right|_i = \sum_{k=1}^L a_k\delta_k\phi_i,$$
where $\delta_k \phi_i = (\phi_{i+k}-\phi_{i-k})/(2 k h)$
the derivative of a product of two functions $f$ and $g$ can be expressed in locally conservative form, namely
as the difference of numerical flux functions $(\hat{F}_{i+1/2}-\hat{F}_{i-1/2})/h$ where $\hat{F}_{i+1/2}$ has the form:
\beq
\hat{F}_{i+1/2} = 2\sum_{k=1}^La_k\sum_{m=0}^{k-1}\mathcal{I}\left(f,g\right)_{i-m,k}\label{eq:Flux}
\eeq
In Eq.~(\ref{eq:Flux}), $\mathcal{I}\left(f,g\right)_{i,k}$ is a suitable two-function two-point interpolation operator.
The \textit{divergence} and \textit{advective} forms of the product $fg$ have the associated interpolation operators,
\beq
\begin{array}{lcl}
\widehat{\dfrac{\partial fg}{\partial x}} &\longrightarrow& \mathcal{I}(f,g)_{i,k}=
(\overline{f,g})_{i,k} \equiv \dfrac{f_{i+k}g_{i+k}+f_{i}g_i}{2},\\
&&\\
\widehat{\left(f\dfrac{\partial g}{\partial x}+g\dfrac{\partial f}{\partial x}\right)} &\longrightarrow& \mathcal{I}\left(f,g\right)_{i,k}=
(\overline{\overline{f,g}})_{i,k} \equiv \dfrac{f_{i+k}g_{i}+f_{i}g_{i+k}}{2}.
\end{array}
\eeq
Any linear combinations of these two forms has a flux function given by the same linear combination of the
corresponding fluxes. The associated interpolations follow the same rule,
\beq
\begin{array}{lcl}
\alpha\widehat{\dfrac{\partial fg}{\partial x}} +
\beta\widehat{\left(f\dfrac{\partial g}{\partial x}+g\dfrac{\partial f}{\partial x}\right)}&\longrightarrow& \mathcal{I}\left(f,g\right)_{i,k}=
\alpha(\overline{f,g})_{i,k} +\beta(\overline{\overline{f,g}})_{i,k}.
\end{array}
\eeq
A particularly simple structure is obtained for the case of $\alpha=\beta=1/2$, for which the
interpolation operator assumes the  simple form
\beq
\begin{array}{lcl}
\dfrac{1}{2}\left[\widehat{\dfrac{\partial fg}{\partial x}} +
\widehat{\left(f\dfrac{\partial g}{\partial x}+g\dfrac{\partial f}{\partial x}\right)}\right]&\longrightarrow& \mathcal{I}\left(f,g\right)_{i,k}=
(\widetilde{f,g})_{i,k}=\dfrac{1}{4}\left(f_{i}+f_{i+k}\right)\left(g_{i}+g_{i+k}\right).
\end{array}
\eeq

In the case of three functions $\rho,u$ and $\phi$ there are five basic ways of expressing the
derivative of the triple product, analogous to the five forms (\ref{eq:CD}--\ref{eq:CL}).
Four of them can be expressed in locally conservative form with associated numerical flux
\beq
\hat{F}_{i+1/2} = 2\sum_{k=1}^La_k\sum_{m=0}^{k-1}\mathcal{I}\left(\rho,u,\phi\right)_{i-m,k}, \label{eq:Flux3}
\eeq
where $\mathcal{I}\left(\rho,u,\phi \right)_{i,k}$ is a suitable three-function two-point interpolation operator.
The list of forms and associated interpolation operators is:
\begin{eqnarray}
\dfrac{\partial \rho u \phi }{\partial x}&\longrightarrow&
\mathcal{I}(\rho,u,\phi)_{i,k}= (\overline{\rho,u,\phi})_{i,k} = \dfrac{(\rho u\phi)_{i+k}+(\rho u\phi)_{i}}{2},\\
\phi\dfrac{\partial \rho u}{\partial x} + \rho u\dfrac{\partial \phi}{\partial x} &\longrightarrow&
\mathcal{I}(\rho,u,\phi)_{i,k}= (\overline{\overline{\rho u,\phi}})_{i,k} = \dfrac{(\rho u)_{i+k}\phi_i+\phi_{i+k}(\rho u)_{i}}{2},\\
u\dfrac{\partial \rho \phi}{\partial x} + \rho \phi\dfrac{\partial u}{\partial x} &\longrightarrow&
\mathcal{I}(\rho,u,\phi)_{i,k}= (\overline{\overline{\rho\phi,u}})_{i,k} = \dfrac{(\rho\phi)_{i+k}u_i+u_{i+k}(\rho\phi)_{i}}{2},\\
\rho \dfrac{\partial u \phi}{\partial x} + \phi u\dfrac{\partial \rho}{\partial x} &\longrightarrow&
\mathcal{I}(\rho,u,\phi)_{i,k}= (\overline{\overline{\phi u,\rho}})_{i,k} = \dfrac{(\phi u)_{i+k}\rho_i+\rho_{i+k}(\phi u)_{i}}{2}.
\end{eqnarray}
Any linear combination of these forms has a flux function whose associated interpolation 
operator $\mathcal{I}$
is a linear combination of the corresponding basic interpolation operators.
Particularly simple structures are obtained in the following cases:
\begin{enumerate}
\item the F form, whose interpolation operator is
\beq
\mathcal{I}(\rho,u,\phi)_{i,k}=
\dfrac{1}{2}\left((\overline{\rho,u,\phi})_{i,k}+ (\overline{\overline{\rho u,\phi}})_{i,k}\right) =
(\widetilde{\rho u,\phi})_{i,k} ;
\eeq
\item the KGP form, whose interpolation operator is
\beq
\mathcal{I}(\rho,u,\phi)_{i,k}=
\dfrac{1}{4}\left((\overline{\rho,u,\phi})_{i,k}+ (\overline{\overline{\rho u,\phi}})_{i,k}+
(\overline{\overline{\rho\phi,u}})_{i,k}+(\overline{\overline{\phi u,\rho}})_{i,k}\right) =
(\widetilde{\rho,u,\phi})_{i,k} ,
\eeq
where $$(\widetilde{\rho,u,\phi})_{i,k}=
\dfrac{1}{8}\left(\rho_{i}+\rho_{i+k}\right)\left(u_{i}+u_{i+k}\right)\left(\phi_{i}+\phi_{i+k}\right);$$
\item the C form, whose interpolation operator is
\beq
\mathcal{I}(\rho,u,\phi)_{i,k}=
\dfrac{1}{2}\left((\overline{\overline{\rho\phi,u}})_{i,k}+(\overline{\overline{\phi u,\rho}})_{i,k}\right) =
(\overline{\overline{\phi|\rho,u}})_{i,k} ,
\eeq
where $$(\overline{\overline{\phi|\rho,u}})_{i,k}=
\dfrac{1}{2}\left(\phi_{i}+\phi_{i+k}\right)(\overline{\overline{\rho,u}})_{i,k}.$$

\end{enumerate}

%

  \bibliographystyle{elsarticle-num-names}

\bibliography{JCP_2018}

\end{document}

%% file: preamble.tex
\newcommand{\beq}{\begin{equation}}
\newcommand{\eeq}{\end{equation}}
\newcommand{\beqs}{\begin{equation*}}
\newcommand{\eeqs}{\end{equation*}}
\newcommand{\bit}{\begin{itemize}}
\newcommand{\eit}{\end{itemize}}
\newcommand{\ben}{\begin{enumerate}}
\newcommand{\een}{\end{enumerate}}

\newcommand{\mC}{\mathcal{C}}
\newcommand{\mQ}{\mathcal{Q}}
\newcommand{\mE}{\mathcal{E}}
\newcommand{\mM}{\mathcal{M}}
\newcommand{\mP}{\mathcal{P}}
\newcommand{\mG}{\mathcal{G}}
\newcommand{\mD}{\mathcal{D}}
\newcommand{\mS}{\mathcal{S}}
\newcommand{\mH}{\mathcal{H}}
\newcommand{\mbE}{\mathbb{E}}

%% file: JCP_2018.bbl
\begin{thebibliography}{21}
\providecommand{\natexlab}[1]{#1}
\providecommand{\url}[1]{\texttt{#1}}
\providecommand{\urlprefix}{URL }
\expandafter\ifx\csname urlstyle\endcsname\relax
  \providecommand{\doi}[1]{doi:\discretionary{}{}{}#1}\else
  \providecommand{\doi}[1]{doi:\discretionary{}{}{}\begingroup
  \urlstyle{rm}\url{#1}\endgroup}\fi
\providecommand{\bibinfo}[2]{#2}

\bibitem[{Phillips(1959)}]{Phillips1959}
\bibinfo{author}{N.~A. Phillips}, \bibinfo{title}{An example of nonlinear
  computational instability}, in: \bibinfo{booktitle}{The {A}tmosphere and the
  {S}ea in {M}otion}, \bibinfo{publisher}{Rockefeller Institute Press, Oxford
  University Press}, \bibinfo{pages}{501 -- 504}, \bibinfo{year}{1959}.

\bibitem[{Lilly(1965)}]{lilly_65}
\bibinfo{author}{D.~K. Lilly}, \bibinfo{title}{On the computational stability
  of numerical solutions of time-dependent non-linear geophysical fluid
  dynamics problems}, \bibinfo{journal}{J. Comput. Phys.} \bibinfo{volume}{93}
  (\bibinfo{year}{1965}) \bibinfo{pages}{11--26}.

\bibitem[{Perot(2011)}]{PerotAR}
\bibinfo{author}{J.~B. Perot}, \bibinfo{title}{Discrete conservation properties
  of unstructured mesh schemes}, \bibinfo{journal}{Annu. Rev. Fluid. Mech.}
  \bibinfo{volume}{43} (\bibinfo{year}{2011}) \bibinfo{pages}{299--318}.

\bibitem[{Jameson(2008)}]{Jameson2008}
\bibinfo{author}{A.~Jameson}, \bibinfo{title}{The Construction of Discretely
  Conservative Finite Volume Schemes that Also Globally Conserve Energy or
  Entropy}, \bibinfo{journal}{J. Sci. Comput.} \bibinfo{volume}{34}
  (\bibinfo{year}{2008}) \bibinfo{pages}{152 -- 187}.

\bibitem[{Kok(2009)}]{Kok2009}
\bibinfo{author}{J.~C. Kok}, \bibinfo{title}{A high-order low-dispersion
  symmetry-preserving finite-volume method for compressible flow on curvilinear
  grids}, \bibinfo{journal}{J. Comput. Phys.} \bibinfo{volume}{228}
  (\bibinfo{year}{2009}) \bibinfo{pages}{6811--6832}.

\bibitem[{Rozema et~al.(2014)Rozema, Kok, Verstappen, and
  Veldman}]{RozemaJT2014}
\bibinfo{author}{W.~Rozema}, \bibinfo{author}{J.~C. Kok},
  \bibinfo{author}{R.~W. C.~P. Verstappen}, \bibinfo{author}{A.~E.~P. Veldman},
  \bibinfo{title}{A symmetry-preserving discretisation and regularisation model
  for compressible flow with application to turbulent channel flow},
  \bibinfo{journal}{J. Turbul.} \bibinfo{volume}{34} (\bibinfo{year}{2014})
  \bibinfo{pages}{386 -- 410}.

\bibitem[{Coppola et~al.(2017)Coppola, Capuano, and
  de~Luca}]{Coppola_AIMETA2017}
\bibinfo{author}{G.~Coppola}, \bibinfo{author}{F.~Capuano},
  \bibinfo{author}{L.~de~Luca}, \bibinfo{title}{Energy-preserving
  discretizations of the {N}avier-{S}tokes equations. Classical and modern
  approaches}, in: \bibinfo{editor}{L.~Ascione}, \bibinfo{editor}{V.~Berardi},
  \bibinfo{editor}{L.~Feo}, \bibinfo{editor}{F.~Fraternali},
  \bibinfo{editor}{A.~M. Tralli} (Eds.), \bibinfo{booktitle}{AIMETA 2017 -
  Proceedings of the 23rd Conference of the Italian Association of Theoretical
  and Applied Mechanics}, vol.~\bibinfo{volume}{3}, \bibinfo{pages}{2284 --
  2310}, \bibinfo{year}{2017}.

\bibitem[{Kravhcenko and Moin(1997)}]{Kravchenko1997}
\bibinfo{author}{A.~G. Kravhcenko}, \bibinfo{author}{P.~Moin},
  \bibinfo{title}{On the effect of numerical errors in large eddy simulations
  of turbulent flows}, \bibinfo{journal}{J. Comput. Phys.}
  \bibinfo{volume}{131} (\bibinfo{year}{1997}) \bibinfo{pages}{310--322}.

\bibitem[{Blaisdell et~al.(1996)Blaisdell, Spyropoulos, and
  Qin}]{Blaisdell1996}
\bibinfo{author}{G.~A. Blaisdell}, \bibinfo{author}{E.~T. Spyropoulos},
  \bibinfo{author}{J.~H. Qin}, \bibinfo{title}{The effect of the formulation of
  nonlinear terms on aliasing errors in spectral methods},
  \bibinfo{journal}{Applied Numerical Mathematics}
  \bibinfo{volume}{21}~(\bibinfo{number}{3}) (\bibinfo{year}{1996})
  \bibinfo{pages}{207 -- 219}.

\bibitem[{Feiereisen et~al.(1981)Feiereisen, Reynolds, and
  Ferziger}]{Feiereisen1981}
\bibinfo{author}{W.~J. Feiereisen}, \bibinfo{author}{W.~C. Reynolds},
  \bibinfo{author}{J.~H. Ferziger}, \bibinfo{title}{Numerical Simulation of
  Compressible, Homogeneous Turbulent Shear Flow}, \bibinfo{type}{Tech. Rep.}
  \bibinfo{number}{TF-13}, \bibinfo{institution}{Stanford University},
  \bibinfo{year}{1981}.

\bibitem[{Morinishi et~al.(1998)Morinishi, Lund, Vasilyev, and
  Moin}]{Morinishi1998}
\bibinfo{author}{Y.~Morinishi}, \bibinfo{author}{T.~S. Lund},
  \bibinfo{author}{O.~V. Vasilyev}, \bibinfo{author}{P.~Moin},
  \bibinfo{title}{Fully conservative higher order finite difference schemes for
  incompressible flows}, \bibinfo{journal}{J. Comput. Phys.}
  \bibinfo{volume}{143} (\bibinfo{year}{1998}) \bibinfo{pages}{90--124}.

\bibitem[{Morinishi(2010)}]{Morinishi2010}
\bibinfo{author}{Y.~Morinishi}, \bibinfo{title}{Skew-symmetric form of
  convective terms and fully conservative finite difference schemes for
  variable density low-Mach number flows}, \bibinfo{journal}{J. Comput. Phys}
  \bibinfo{volume}{229} (\bibinfo{year}{2010}) \bibinfo{pages}{276--300}.

\bibitem[{Kennedy and Gruber(2008)}]{Kennedy2008}
\bibinfo{author}{C.~A. Kennedy}, \bibinfo{author}{A.~Gruber},
  \bibinfo{title}{Reduced aliasing formulations of the convective terms within
  the {N}avier-{S}tokes equations for a compressible fluid},
  \bibinfo{journal}{J. Comput. Phys.} \bibinfo{volume}{227}
  (\bibinfo{year}{2008}) \bibinfo{pages}{1676--1700}.

\bibitem[{Johnsen et~al.(2010)Johnsen, Larsson, Bhagatwala, Cabot, Moin, Olson,
  Rawat, Shankar, Sj{\"o}green, Yee, Zhong, and Lele}]{JohnsenJCP2010}
\bibinfo{author}{E.~Johnsen}, \bibinfo{author}{J.~Larsson},
  \bibinfo{author}{A.~V. Bhagatwala}, \bibinfo{author}{W.~H. Cabot},
  \bibinfo{author}{P.~Moin}, \bibinfo{author}{B.~J. Olson},
  \bibinfo{author}{P.~S. Rawat}, \bibinfo{author}{S.~K. Shankar},
  \bibinfo{author}{B.~Sj{\"o}green}, \bibinfo{author}{H.~Yee},
  \bibinfo{author}{X.~Zhong}, \bibinfo{author}{S.~K. Lele},
  \bibinfo{title}{Assessment of high-resolution methods for numerical
  simulations of compressible turbulence with shock waves},
  \bibinfo{journal}{J. Comput. Phys.} \bibinfo{volume}{229}
  (\bibinfo{year}{2010}) \bibinfo{pages}{1213 -- 1237}.

\bibitem[{Pirozzoli(2010)}]{Pirozzoli2010}
\bibinfo{author}{S.~Pirozzoli}, \bibinfo{title}{Generalized conservative
  approximations of split convective derivative operators},
  \bibinfo{journal}{J. Comput. Phys.}
  \bibinfo{volume}{229}~(\bibinfo{number}{19}) (\bibinfo{year}{2010})
  \bibinfo{pages}{7180 -- 7190}.

\bibitem[{Honein and Moin(2004)}]{Honein2004}
\bibinfo{author}{A.~E. Honein}, \bibinfo{author}{P.~Moin},
  \bibinfo{title}{Higher entropy conservation and numerical stability of
  compressible turbulence simulations}, \bibinfo{journal}{J. Comput. Phys.}
  \bibinfo{volume}{201} (\bibinfo{year}{2004}) \bibinfo{pages}{531--545}.

\bibitem[{Mansour et~al.(1979)Mansour, Moin, Reynolds, and
  Ferziger}]{Mansour1979}
\bibinfo{author}{N.~N. Mansour}, \bibinfo{author}{P.~Moin},
  \bibinfo{author}{W.~C. Reynolds}, \bibinfo{author}{J.~H. Ferziger},
  \bibinfo{title}{Improved methods for large eddy simulations of turbulence},
  \bibinfo{journal}{Turb. Shear Flows} \bibinfo{volume}{1}
  (\bibinfo{year}{1979}) \bibinfo{pages}{386--401}.

\bibitem[{Capuano et~al.(2017)Capuano, Coppola, R{\'a}ndez, and
  de~Luca}]{capuano2017explicit}
\bibinfo{author}{F.~Capuano}, \bibinfo{author}{G.~Coppola},
  \bibinfo{author}{L.~R{\'a}ndez}, \bibinfo{author}{L.~de~Luca},
  \bibinfo{title}{Explicit {R}unge-{K}utta schemes for incompressible flow with
  improved energy-conservation properties}, \bibinfo{journal}{J. Comput. Phys.}
  \bibinfo{volume}{328} (\bibinfo{year}{2017}) \bibinfo{pages}{86--94}.

\bibitem[{Verstappen and Veldman(2003)}]{Verstappen2003}
\bibinfo{author}{R.~W. C.~P. Verstappen}, \bibinfo{author}{A.~E.~P. Veldman},
  \bibinfo{title}{Symmetry--preserving discretization of turbulent flow},
  \bibinfo{journal}{J. Comput. Phys.} \bibinfo{volume}{187}
  (\bibinfo{year}{2003}) \bibinfo{pages}{343--368}.

\bibitem[{Capuano et~al.(2015)Capuano, Coppola, Balarac, and
  de~Luca}]{Capuano2015b}
\bibinfo{author}{F.~Capuano}, \bibinfo{author}{G.~Coppola},
  \bibinfo{author}{G.~Balarac}, \bibinfo{author}{L.~de~Luca},
  \bibinfo{title}{Energy preserving turbulent simulations at a reduced
  computational cost}, \bibinfo{journal}{J. Comput. Phys.}
  \bibinfo{volume}{298} (\bibinfo{year}{2015}) \bibinfo{pages}{480--494}.

\bibitem[{Shu and Osher(1989)}]{ShuOsher1989}
\bibinfo{author}{C.-W. Shu}, \bibinfo{author}{S.~Osher},
  \bibinfo{title}{Efficient implementation of essentially non-oscillatory
  shock-capturing schemes, II}, \bibinfo{journal}{J. Comput. Phys.}
  \bibinfo{volume}{83} (\bibinfo{year}{1989}) \bibinfo{pages}{32 -- 78}.

\end{thebibliography}
